\documentclass[pra,twocolumn,aps,superscriptaddress,floatfix,tightenlines]{revtex4-2}
\usepackage{amsmath}
\usepackage{amssymb}
\usepackage{amsfonts}
\usepackage{epsfig}
\usepackage{bm}
\usepackage{times}
\usepackage[T1]{fontenc}
\usepackage[colorlinks,linkcolor=blue,anchorcolor=blue,citecolor=blue,urlcolor=blue]{hyperref}
\usepackage{braket}
\usepackage{graphicx}
\usepackage{setspace}
\usepackage{subfigure}
\usepackage{enumerate}
\begin{document}

    \title{Recovery of damaged information via scrambling in indefinite causal order}
    
    \author{Tian-Ren Jin}
        \affiliation{Institute of Physics, Chinese Academy of Sciences, Beijing 100190, China}
        \affiliation{School of Physical Sciences, University of Chinese Academy of Sciences, Beijing 100049, China}

    \author{Tian-Ming Li}
        \affiliation{Institute of Physics, Chinese Academy of Sciences, Beijing 100190, China}
        \affiliation{School of Physical Sciences, University of Chinese Academy of Sciences, Beijing 100049, China}

    \author{Zheng-An Wang}
        \affiliation{Beijing Academy of Quantum Information Sciences, Beijing 100193, China}
        \affiliation{Hefei National Laboratory, Hefei 230088, China}

    \author{Kai Xu}
        \affiliation{Institute of Physics, Chinese Academy of Sciences, Beijing 100190, China}
        \affiliation{School of Physical Sciences, University of Chinese Academy of Sciences, Beijing 100049, China}
        \affiliation{Beijing Academy of Quantum Information Sciences, Beijing 100193, China}
        \affiliation{Hefei National Laboratory, Hefei 230088, China}
        \affiliation{Songshan Lake Materials Laboratory, Dongguan 523808, China}
        \affiliation{CAS Center for Excellence in Topological Quantum Computation, UCAS, Beijing 100190, China}

    \author{Yu-Ran Zhang}
        \email{yuranzhang@scut.edu.cn}
        \affiliation{School of Physics and Optoelectronics, South China University of Technology, Guangzhou 510640, China}

    \author{Heng Fan}
        \email{hfan@iphy.ac.cn}
        \affiliation{Institute of Physics, Chinese Academy of Sciences, Beijing 100190, China}
        \affiliation{School of Physical Sciences, University of Chinese Academy of Sciences, Beijing 100049, China}
        \affiliation{Beijing Academy of Quantum Information Sciences, Beijing 100193, China}
        \affiliation{Hefei National Laboratory, Hefei 230088, China}
        \affiliation{Songshan Lake Materials Laboratory, Dongguan 523808, China}
        \affiliation{CAS Center for Excellence in Topological Quantum Computation, UCAS, Beijing 100190, China}

    \begin{abstract}
        Scrambling prevents the access to local information with local operators and therefore can be used to protect
        quantum information from damage caused by local perturbations.
        Even though partial quantum information can be recovered if the type of the damage is known,
        the initial target state cannot be completely recovered,
        because the obtained state is a mixture of the initial state and a maximally mixed state.
        Here, we demonstrate an improved scheme to recover
        damaged quantum information via scrambling in indefinite causal order.
        We show that scheme with  indefinite causal order can record information of the damage and distill the initial state from the damaged state simultaneously.
        It allows us to retrieve initial information versus any damage. 
        Moreover, by iterating the schemes, the initial quantum state can be completely recovered.
        In addition, we experimentally demonstrate our schemes on the
        cloud-based quantum computer, named as \emph{Quafu}. Our work proposes a feasible
        scheme to protect whole quantum information from damage, which is also
        compatible with other techniques such as quantum error corrections and
        entanglement purification protocols.
        We expect that our scheme will be useful in the both
        quantum information recovery from the damage and system’s
        bench-marking.%
    \end{abstract}

    \maketitle

    \section{Introduction}

        Quantum scrambling in a chaotic system can be described as the state being randomized with respect to the Haar measure over the entire Hilbert space~\cite{sekino2008fast}.
        In the view of thermalization, different initial states cannot be distinguished with local measurements of the system~\cite{lashkari2013towards}.
        This effect was first studied in the dynamics of black holes~\cite{Hayden_2007,sekino2008fast, lashkari2013towards,shenker2014black,maldacena2016bound} and has been attracting growing attention for its close relation to quantum chaos and thermalization in isolated quantum many-body systems~\cite{PhysRevA.43.2046,PhysRevE.50.888,Srednicki_1999,rigol2008thermalization}.
        Recently, it has been realized in a variety of systems, including nuclear magnetic resonance (NMR)~\cite{PhysRevLett.124.250601}, trapped ions~\cite{PhysRevLett.124.240505,PhysRevLett.128.140601,landsman2019verified}, and superconducting~\cite{mi2021information,mi2021information,PhysRevLett.129.160602,braumuller2022probing,PhysRevX.11.021010}.
        Moreover, it relates to quantum information theory~\cite{RevModPhys.91.025001,Gour_2008}.

        Black holes were once considered as monsters only devouring everything. Then, Hawking radiation reveals that black holes will completely evaporate internal information eventually.
        By assuming that black holes just process and do not destroy information, it has been shown that information cast into black hole can be recovered, by collecting the Hawking radiation more bits of information than its scrambled~\cite{Hayden_2007}.
        To comply with the quantum no-cloning theorem, the scrambling time should be bounded by an order of $\sim S^{2/D}$, where $S$ and $D$ are the entropy and the dimension of system, respectively~\cite{sekino2008fast}. Black holes are the fastest scramblers with an infinite dimension, namely, saturating the bound with an order of $\sim \log S$.
        This conjecture is also supported by other works in Refs.~\cite{lashkari2013towards,shenker2014black,maldacena2016bound}.

        Since quantum information cast into the black hole will be emitted after scrambling, this mechanism can be used to recover information.
        In this Hayden-Preskill black hole thought experiment~\cite{Hayden_2007}, the quantum information theory is used to show that with the early Hawking radiation and the radiation after scrambling, initial information can be recovered.
        One of the specific recipes is a teleportation-based decoding protocol~\cite{yoshida2017efficient,vermersch2019probing,landsman2019verified}, which is similar to a traversable wormhole.
        With the knowledge of the black hole dynamics, information can be retrieved through a quantum memory, pre-entangled with the black hole, by collecting the Hawking radiation.
        However, this protocol actually collects all initial information, both emitted by the Hawking radiation and left in the black hole, which unrealistically requires considerable knowledge of the black hole system.

        A recent scheme, proposed by B. Yan \emph{et al.}~\cite{PhysRevLett.125.040605}, reconstructs initial information without a complete knowledge of the radiation.
        In details, Alice encodes information by scrambling it with the black hole and can recover information through the backward evolution of the black hole dynamics, where the correlations between radiation and the black hole may be damaged in some degree by the attacker, Bob.
        This scheme helps to protect and retrieve quantum information against the damage.
        However, there are still two limitations in this protocol:
        ($i$) A complete retrieval of initial information relies on the knowledge of the type of the damage.
        ($ii$) The initial quantum state can only be partially obtained as a mixture of the initial state with
        the maximally mixed state.

        In this paper, we propose an improved scheme for recovering damaged quantum information to cope with the problems by applying the scrambling recovery scheme with the quantum effect of the indefinite causal order (ICO)~\cite{oreshkov2012correlaciones}.
        Here, auxiliary qubits are used to record the type of damage, with which the first problem is resolved.  These auxiliary qubits help to improve the fidelity of recovered state with respect to the initial state, which weaken the requirement of the fidelity of encoding when applying quantum error correction codes (QECCs).
        In particular, for the damage caused by one-qubit projective measurement, there is a measurement-direction-dependent (MDD) scheme to recover the initial state completely.
        The scheme can also be iterated, which has the potential to recover the initial state completely from all damage. Our methods to recover the damaged information via scrambling is different from the QECCs and have lower requirement of the fidelity than QECCs, and thus can pre-recover the damaged state over the required fidelity of the QECCs.
        Our work provides a method to record the type of damage and to distill the damaged quantum state,
        which allows to retrieve initial information versus any damage.

        This paper is organized as follows.
        In Sec.~\ref{sec: origin}, information scrambling and the original protocol proposed in Ref.~\cite{PhysRevLett.125.040605} are reviewed.
        The improved schemes are discussed in Sec.~\ref{sec: qubit}, and the iterative schemes are discussed in Sec.~\ref{sec: qudit}.
        In Sec.~\ref{sec: simulation}, the quantum simulations of several schemes are performed on the \emph{Quafu}
         on-cloud quantum computer.
        The conclusion and discussion are given in Sec.~\ref{sec: conclusion}.

    \section{Preliminaries} \label{sec: origin}
        \subsection{Scrambling and Out-of-Time Correlations}
            Information scrambling renders that any two different states will be locally indistinguishable~\cite{lashkari2013towards}, due to the chaotic dynamical evolution.
            It spreads local information and generates entanglement between different particles,
            which prevents the access to initial information from local measurement.
            Two states $\rho_1$ and $\rho_2$ are locally indistinguishable, if the reduced matrix
            $\rho_1^S(t)$ and $\rho_2^S(t)$ of the two states on a sufficient small subsystem $S$
            of a $N$-particle system can be arbitrarily close when time $t$ is long enough.
            More precisely, for any given small parameter $\delta$, there exists a time $t_0$ and
            a dimensionless parameter $\epsilon$, when $t > t_0$ and the subsystem size  $|S| < \epsilon N$, $\Vert \rho_1^S(t) - \rho_2^S(t) \Vert < \delta$ with respect to some norm~\cite{lashkari2013towards}.
            This property allows thermalization occurring on a subsystem while the whole system is still out of equilibrium, which releases the contradiction between thermal equilibrium and quantum invertibility~\cite{lashkari2013towards,mi2021information}.

            For operator scrambling, we define a direct product operator as
            $\hat{O} = \bigotimes_{i = 1}^N \hat{o}_i$.
            First, we consider a local operator initially acting only on the first particle as
            $\hat{O}_{H}(0)=\hat{o}_1 \otimes \hat{\mathbb{I}}_{\{2,\dots,N\}}$ and then 
            evolving chaotically as
            \begin{equation}\label{eq1}
                 \hat{O}_{H}(t)  = \hat{U}^{\dagger}(t) \hat{O} \hat{U}(t) = \sum_{l} s_l \hat{O}^{(l)},
             \end{equation}
            where $\hat{O}^{(l)}$ is a $l$-weight term acting  on  $l$ particles.
            As the time increases, larger weight terms will be involved in the summation of $\hat{O}_{H}(t)$~\cite{mi2021information}.
            This effect can be detected with the OTOCs
            \begin{equation}
                C(t) = \langle{\hat{O}_{H}^{\dagger}(t) \hat{W}^{\dagger} \hat{O}_{H}(t) \hat{W}}\rangle,
            \end{equation}
            where $\hat{W}$ is assumed to only act on the $N$-th particle, and without loss of generality,
            $\hat{O}$ and $\hat{W}$ are assumed to be unitary.
            At an early time,   $\hat{O}_{H}(t)$, acts trivially on the $N$-th particle.
            Thus, they commute with each other, and the OTOCs $C(t)$ approach to 1.
            However, as the time increases, terms $\hat{O}^{(l)}$ with a larger weight will be involved in the summation (\ref{eq1}),
            which will act non-trivially on the $N$-th particle. Then, $\hat{O}_{H}(t)$ does not commute with $\hat{W}$,
            leading to a decay of $C(t)$ from 1.

        \subsection{Original Recovery Protocol via Scrambling}
            In the original recovery protocol via scrambling as proposed in
            Ref.~\cite{PhysRevLett.125.040605}, Alice encodes the state of the target system
            into a black hole, by performing a chaotic evolution of the whole system to fast
            scramble local information.
            Then, the target system is exposed to the attacker Bob, who performs some perturbation
            $\Lambda$ on it, e.g., a projective measurement.
            The target system should be interpreted as the system exposed to the perturbation $\Lambda$, while the bath (black hole), as a part of the whole system, is not exposed to perturbation.
            Note that it is possible that the whole system is exposed to perturbation, thus the bath is a virtual one.
            After the attack, Alice decodes the target system by backward evolving the whole system.
            Then, she can retrieve the initial quantum state partially.

\begin{figure}[t]
                \centering
                \includegraphics[width=0.4\textwidth]{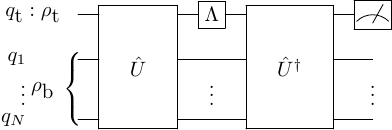}
                \caption{Quantum circuit (using qubits as an example) for the original protocol for recovering damaged information via scrambling, where $q_{\textrm{t}}$ is target system, $q_1, \dots, q_N$ denote the bath (black hole).}
                \label{fig: scrambling}
            \end{figure}

            This procedure is described by the quantum circuit in Fig.~\ref{fig: scrambling}.
            The initial state is assumed as
            $\rho_{\textrm{tb}} ^{\textrm{in}}=\rho_{\textrm{t}}\otimes\rho_{\textrm{b}}$, with $\rho_{\textrm{t}}$
            and $\rho_{\textrm{b}}$ being the density matrices of the target system (t) and the bath (b),
            respectively.
            In the Hayden-Preskill problem, the black hole is assumed to be a fast scrambler~\cite{Hayden_2007,sekino2008fast}.
            Therefore, the black hole is characterized by the chaotic dynamics $\mathcal{U}$ as shown in Fig.~\ref{fig: scrambling}.
            Moreover, since the state of the black hole $\rho_{\textrm{b}}$ is unknown to the observer, i.e. Alice and Bob, out of the horizon, so $\rho_{\textrm{b}}$ should be assumed as an arbitrary possible state. 

            For a long-time evolution, the chaotic dynamics can be described as random unitary evolutions
            uniformly under the Haar measure on the unitary transformation group~\cite{shenker2014black,PhysRevLett.125.040605,PhysRevLett.129.050602}.
            This protocol can be understood as a twirling channel in the quantum reference frame theory
            \cite{RevModPhys.91.025001}, i.e., the perturbation $\Lambda$ (quantum operation) is twirled by the random unitary
            transformations $\mathcal{U}(\cdot) = \hat{U} (\cdot) \hat{U}^{\dagger}$, and the output state
            of the total system as
            \begin{align}
                \rho_{\mathrm{tb}}^{\mathrm{out}} = \Lambda_{\mathrm{twirl}}(\rho^{\mathrm{in}}_{\mathrm{tb}}) & = \int \!\!\mathrm{d}\mathcal{U} \;  \mathcal{U} \circ \Lambda \circ \mathcal{U}^{\dagger}(\rho^{\mathrm{in}}_{\mathrm{tb}} ) \\
                & = \int \!\!\mathrm{d}\hat{U}\; \sum_k \hat{U} \hat{M}_k^{\dagger} \hat{U}^{\dagger} \rho^{\mathrm{in}}_{\mathrm{tb}}  \hat{U} \hat{M}_k \hat{U}^{\dagger},\nonumber
            \end{align}
            where $\hat{M}_k$ is the Kraus operators of the completely positive and trace preserving (CPTP) perturbation $\Lambda(\cdot) = \sum_k \hat{M}_k^{\dagger}(\cdot)\hat{M}_k$ with the identity $\sum_k\hat{M}_k\hat{M}_k^{\dagger}=\hat{\mathbb{I}}$.
            Using Weingarten's function~\cite{10.1155/S107379280320917X,collins2006integration}, we have
            \begin{align}
                &(D^2_{\textrm{tb}} - 1)\int\! d\hat{U}\; \hat{U}_{m_1 n_1} \hat{U}_{k_1 l_1}^{\ast} \hat{U}_{m_2 n_2} \hat{U}_{k_2 l_2}^{\ast} \nonumber\\
                &~~=\delta_{m_1 k_1} \delta_{m_2 k_2} \delta_{n_1 l_1} \delta_{n_2 l_2} + \delta_{m_1 k_2} \delta_{m_2 k_1} \delta_{n_1 l_2} \delta_{n_2 l_1}                 \label{eq: Weingarten}\\
                &~~~~~~- (\delta_{m_1 k_1} \delta_{m_2 k_2} \delta_{n_1 l_2} \delta_{n_2 l_1} + \delta_{m_1 k_2} \delta_{m_2 k_1} \delta_{n_1 l_1} \delta_{n_1 l_1})/D_{\textrm{tb}},\nonumber
                \end{align}
            where $\delta_{mn}$ is Kronecker delta function, and $D_{\textrm{tb}}$ is the dimension of the composited system,
            combining the target system (t) and the black hole (b).
            Then, we have
            \begin{equation} 
                \rho_{\textrm{tb}}^{\textrm{out}}=\Lambda_{\mathrm{twirl}}(\rho_{\textrm{tb}} ^{\textrm{in}}) = p\rho_{\textrm{tb}}^{\textrm{in}} + {(1-p )}\hat{\mathbb{I}}_{\textrm{tb}}/{D}_{\textrm{tb}},
            \end{equation}
            where the recovery rate
            \begin{equation} \label{eq: recovery}
                p = \frac{\sum_k |\mathrm{Tr}\hat{M}_k|^2 - 1}{D_{\textrm{tb}}^2 - 1}
            \end{equation}
            is determined by the type of the perturbation $\Lambda$.
            The output state of the target system can be written as
            \begin{equation} \label{eq: twirling}
                \rho^{\textrm{out}}_{\textrm{t}}=p\rho_{\textrm{t}} + {(1-p)}\hat{\mathbb{I}}_{\textrm{t}} /{D}_{\textrm{t}},
            \end{equation}
            with $D_{\textrm{t}}$ being the dimension of the target system.
            By performing quantum state tomography (QST) on the target system,
            Alice can retrieve quantum information of $\rho$ from the damage with the knowledge of $p$.
            For example, consider a measurement on target qubit (${D}_{\textrm{t}}=2$) with $\hat{M}_{\pm \bm{k}} = ({\hat{\mathbb{I}} \pm \bm{k} \cdot \bm{\sigma}})/{2}$, where $\bm{k} = (k_x, k_y, k_z)$ is the measurement direction, and $\bm{\sigma} = (\hat{\sigma}^x, \hat{\sigma}^y, \hat{\sigma}^z)$ with $\hat{\sigma}^{x,y,z}$ being Pauli operators.
            For $D_{\textrm{tb}} \rightarrow \infty$ (e.g., a black hole), we have
            \begin{equation}
            p = \frac{D_{\textrm{tb}}^2/2 - 1}{D_{\textrm{tb}}^2 - 1} \rightarrow \frac{1}{2},
            \end{equation}
            indicating that Alice can recover initial quantum information with
            a recover rate $p=1/2$.

            In addition, note that the recovery rate $p$ is related to the average fidelity of the perturbed state with initial states from the Haar ensemble.
            The average fidelity is expressed as
            \begin{align}
                \bar{F}(\Lambda) & = \int \!\!d\mathcal{U} \; \mathrm{Tr}[\mathcal{U}(\rho_{\textrm{tb}} ^{\textrm{in}}) \Lambda \circ \mathcal{U}(\rho_{\textrm{tb}} ^{\textrm{in}})] \nonumber \\
                & = \mathrm{Tr}\left[\rho_{\textrm{tb}} ^{\textrm{in}} \int \!\!d\mathcal{U} \; \mathcal{U}^{\dagger} \circ \Lambda \circ \mathcal{U} (\rho_{\textrm{tb}} ^{\textrm{in}})\right]\nonumber\\
                & = \mathrm{Tr}\left[\rho_{\textrm{tb}} ^{\textrm{in}}\left(p\rho_{\textrm{tb}} ^{\textrm{in}} + {(1-p )}\hat{\mathbb{I}}_{\textrm{tb}}/{D}_{\textrm{tb}},\right)\right] \nonumber\\
                & = (1 - 1/{D}_{\textrm{tb}}) p + 1/{D}_{\textrm{tb}},
            \end{align}
            where $\rho_{\textrm{tb}} ^{\textrm{in}}$ is chosen as a pure state.

        \subsection{Limitations of the Original Recovery Protocol}
            Before going ahead, we make some assumptions of our schemes in the black hole story.
            First, we cannot monitor or detect the state of black hole directly.
            Second, we do not know quantum information that we want to protect.
            Thus, useful schemes should work for arbitrary initial state of the composited system.

            Then, we discuss the problems mentioned in the Introduction.
            First, $p$ cannot be directly obtained from the QST measurement on the output state,
            when $\rho$ is a mixed state.
            Thus, we need to record the recovery rate $p$ given different kinds of
            damage, which cannot be realized in the original protocol.
            Second, the output state (\ref{eq: twirling}) is a mixture of $\rho_{\textrm{t}}$ with a probability $p$ and $\hat{\mathbb{I}}_{\textrm{t}}/D_{\textrm{t}}$ with $(1-p)$.

            Moreover, when the target system has a ${D}_{\textrm{t}}>2$ dimension  (e.g., a qudit with ${D}_{\textrm{t}}=4$), and the perturbation is assumed to be a projective measurement, the probability is calculated as
            \begin{equation}
                p = \frac{\sum_k |\mathrm{Tr}\hat{M}_k|^2 - 1}{D_{\textrm{tb}}^2 - 1} = \frac{D_{\textrm{tb}}^2/{D}_{\textrm{t}} - 1}{{D}_{\textrm{t}}^2 - 1} \rightarrow \frac{1}{{D}_{\textrm{t}}},
            \end{equation}
            which means that much less initial information can be recovered using the
            original protocol in Ref.~\cite{PhysRevLett.125.040605}. Thus, for a target system with a large dimension, we will need to distill information of the initial state $\rho_{\textrm{t}}$ from the mixture (\ref{eq: twirling}) to overcome this problem.

        \subsection{Recovery Assisted by Quantum Error Correction Codes} 

            As the twirlling perturbation $\Lambda_{\mathrm{twirl}}$ behaves as a depolarizing error on the target system,
            QECCs can be used to protect information from attack.
            Here, we discuss whether QECCs can help to recover information in the viewpoint of the entanglement purification protocol (EPP), since the EPP with one-way classical channels is equivalent to the QECC~\cite{PhysRevA.54.3824,Dur2007aa}.
            Moreover, we aim to distill the initial states from the recovered mixed states.
            It is similar to the EPP, where the entangled pure states are distilled from entangled mixed states.

            The simplest way to demonstrate the equivalence between QECC and one-way EPP is to use the Choi-Jamiołkowsky (CJ) isomorphism, which shows the duality between the quantum states and the quantum operations~\cite{JAMIOLKOWSKI1972275,CHOI1975285}.
            For any CPTP map $\mathcal{E}: A \rightarrow B$ from a system $A$ to a system $B$, there corresponds to a Choi matrix of the composited system $AB$, defined as
            \begin{equation}
                J_{\mathcal{E}}^{AB} = \mathcal{I} \otimes \mathcal{E}(\ket{\phi^+} \bra{\phi^+}),
            \end{equation}
            where $\ket{\phi^+} = \sum_{j=1}^{d_A} \ket{j}_A \ket{j}_{A'}$ is the maximal entangled state of system $A$ and its copy $A'$ up to the normalization.
            The Choi matrix $J_{\mathcal{E}}^{AB}$ is a state of the system $AB$ up to normalization, uniquely corresponding to the quantum operation $\mathcal{E}$.
            The output state of the map $\mathcal{E}$ applied on the state $\rho$ can be represented by the Choi matrix $J_{\mathcal{E}}^{AB}$ as~\cite{RevModPhys.91.025001}
            \begin{equation}
                \mathcal{E}(\rho) = \mathrm{Tr}_A[J_{\mathcal{E}}^{AB} (\rho^T \otimes \hat{\mathbb{I}}_B)].
            \end{equation}
            This expression establishes the CJ isomorphism between maps $\mathcal{E}: A \rightarrow B$, and the density matrix $J_{\mathcal{E}}^{AB}$ of the composited system $AB$ up to the normalization, where $A$ is the input system and $B$ is the output system.

            Thus, in the qubits system, the Bell state $\ket{\Phi^+} = (\ket{00} + \ket{11})/{\sqrt{2}}$ corresponds to the identical channel $\mathcal{I}$, which agrees with the teleportation protocol, and the non-maximal entangled states correspond to  some error channels $\mathcal{E}$.
            In this circumstance, EPP's consuming some copies of non-maximal entangled state to distill maximal entangled state is similar as the fact that  QECC's encoding logic qubit into several physical qubits to defend errors.
            However, for the causality of time, the input state should not be influenced by the output state, thus the EPP, with only one-way classical channel, is equivalent to QECC.
            One can also construct an EPP from a QECC by applying the transformation $\hat{U}_{E}^{T}$ of encoding operator $\hat{U}_{E}$ on the input system $A$, and the decoding operator $\hat{U}_{D}$ on the output system $B$~\cite{Dur2007aa}.

            Here, the Choi matrix of scrambling $\Lambda_{\mathrm{t}}$ corresponds to the entangled state
            \begin{equation}
                \rho_{\textrm{W}}(p) = p \ket{\Phi^+} \bra{\Phi^+} + (1 - p) {\hat{\mathbb{I}}}/{4},
            \end{equation}
            which is called Werner state~\cite{Dur2007aa}, and the fidelity between the Werner state, corresponding to the twirling perturbation of one-qubit projective measurement, and the Bell state $\ket{\Phi^+}$, corresponding to the identity operation lies in $$F_1 = {1}/{2} \leq F = ({1 + 3 p})/{4} < F_{\infty} = {5}/{8}.$$
            To distill the initial state in recovered state is thus equivalent to distill the Bell state from the Werner state corresponding to the twirling perturbation.
            There are lots of purification protocols designed for the Werner state.
            For the $2 \rightarrow 1$ protocol, BBPSSW \cite{PhysRevLett.76.722} and DEJMPS \cite{PhysRevLett.77.2818} protocols require the fidelity $F>{1}/{2}$ of $\ket{\Phi^+}$ with the Werner state, while for $M \rightarrow N$ protocols, breeding protocols works only for $F \gtrapprox 0.81$ \cite{PhysRevLett.76.722,Dur2007aa}.
            When the encoding and decoding operations are not ideal, the minimal required fidelity will increase while the maximal reachable fidelity will decrease, and at some threshold of error rate, the protocol will fail.
            The BBPSSW and DEJMPS protocols cannot distill the state $\ket{\Phi^+}$, if the two-qubit gate fidelities are less that about $0.975$ and $0.965$, and will totally fail if two-qubit gate fidelities are less than $0.965$ and $0.945$ \cite{PhysRevA.59.169}.
            For a target system with a higher dimension, the minimal required fidelity of QECCs or EPPs is not satisfied even for ideal encoding.
            Therefore, we need to find other methods to increase the fidelity $F = ({1 + 3 p})/{4}$ or equivalently the recovery rate $p$ of the recovered state for the scrambling recovery of the damaged information.

    \section{Improved Scrambling Recovery Schemes} \label{sec: qubit}
        In this section, we discuss two improved schemes and their combination.
        In the first scheme,
        the recovery rate $p$, depending on the type of unknown damage,
        is recorded with an additional qubit.
        This allows for retrieving all initial information from any damage by performing
        QST on the target system.
        Meanwhile, the auxiliary qubit also helps to distill the initial state from the recovered mixed state, which weakens the requirement of two-qubit gate fidelity when applying QECCs.
        Second, for the projective measurement as the perturbation, it is possible to recover the initial state with a considerably high accuracy.
        Since the successful probabilities of these schemes are not less than $1/2$, all the advantages mentioned above have practical potential applications.

        It is worth emphasizing that information preserved in the recovered state, determined by the perturbation, cannot be increased by the following schemes.
        These schemes distill initial states from the recovered mixed states of scrambling recovery protocol.
        The effectiveness of the distillation is described by ``yield''~\cite{Dur2007aa}, the ratio $Y_{F,\mathcal{P}} = n/N$ of $n$ distilled states with given fidelity $F$ by a protocol $\mathcal{P}$ to $N$ recovered mixed states consumed.
        It is obvious that the higher the fidelity $F$ of distilled state, the lower the yield $Y_{F,\mathcal{P}}$.
        The yields are quantified by the successful probabilities, denoted as $p_{+}$ or $p_{++}$, in the following.

    \subsection{Indefinite Causal Order Scheme of Scrambling Recovery}

            \begin{figure}[t]
                \centering
                \includegraphics[width=0.49\textwidth]{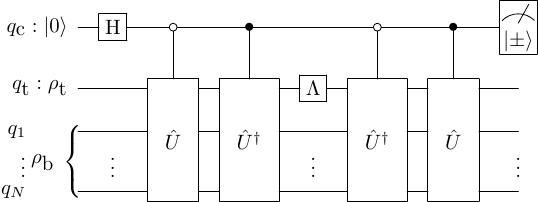}
                \caption{Quantum circuit for the indefinite causal order (ICO) scheme for scrambling recovery of information.}
                \label{fig: ico}
            \end{figure}
            Indefinite causal order (ICO) is a quantum superposition of different sequences of events with different order, which emerges as quantum effect \cite{oreshkov2012correlaciones}.
            Two identical fully depolarizing channels or thermalizing channels performed in ICO will preserve the information of the initial state, which cannot be protected in the classic causal order \cite{PhysRevLett.120.120502,PhysRevLett.125.070603}.
            Therefore, by performing information scrambling in ICO, we can recover more information than the
            original protocol \cite{PhysRevLett.125.040605}.
            The circuit of the quantum-SWITCH of scrambling is shown Fig.~\ref{fig: ico}.

            Here, $q_{\textrm{c}}$ is a control qubit, $\hat{U}$ denotes the random unitary operator, and $\Lambda$ is the perturbation channel.
            The quantum-SWITCH channel is written as
            \begin{equation}
                \mathcal{V}(\rho_{\textrm{ctb}}) = \hat{V} \rho_{\textrm{ctb}}\hat{V}^{\dagger}, \end{equation}
                where
                \begin{equation}
             \hat{V} = {|0\rangle}_{\textrm{c}} {\langle0|} \otimes \hat{U} + {|1\rangle}_{\textrm{c}}{\langle1|} \otimes \hat{U}^{\dagger}.
             \end{equation}
            The quantum twirling channel of the perturbation $\Lambda(\cdot) = \sum_{k} \hat{M}_k^{\dagger} (\cdot) \hat{M}_k$ is
            \begin{equation}
                \Lambda_{\mathrm{twirl}}(\rho_{\textrm{ctb}}) = \mathcal{V}\circ\Lambda\circ\mathcal{V}^{-1}(\rho_{\textrm{ctb}}) = \sum_k \hat{W}_k^{\dagger} \rho _{\textrm{ctb}}\hat{W}_k,
            \end{equation}
            with
            \begin{equation}
                \hat{W}_k =  |0\rangle_{\textrm{c}}  \langle0| \otimes \hat{U} \hat{M}_k \hat{U}^{\dagger} + {|1\rangle}_{\textrm{c}} \langle{1}| \otimes \hat{U}^{\dagger} \hat{M}_k \hat{U} .
            \end{equation}
            Given the input state $\rho^{\mathrm{in}}_{\textrm{ctb}} = |+\rangle_{\textrm{c}}\langle+| \otimes \rho_{\textrm{tb}}^{\textrm{in}}$, the output of the quantum twirling channel is
            \begin{widetext}
                \begin{align}
                        \rho^{\mathrm{out}}_{\textrm{ctb}}=\Lambda_{\mathrm{twirl}} (\rho^{\mathrm{in}}_{\textrm{ctb}}) & = \frac{1}{2} \sum_k \left[|{0}\rangle_{\textrm{c}} \langle{0} |\otimes \hat{U} \hat{M}_k^{\dagger} \hat{U}^{\dagger} \rho_{\textrm{tb}}^{\textrm{in}} \hat{U} \hat{M}_k \hat{U}^{\dagger}
                        + |{0}\rangle_{\textrm{c}} \langle{1} |\otimes \hat{U} \hat{M}_k^{\dagger} \hat{U}^{\dagger} \rho_{\textrm{tb}}^{\textrm{in}} \hat{U}^{\dagger} \hat{M}_k \hat{U}\right. \\
                        & \quad + \left. |{1}\rangle_{\textrm{c}} \langle{0}| \otimes U^{\dagger} M_k U^{\dagger} \rho_{\textrm{tb}}^{\textrm{in}} U M_k U^{\dagger}
                        + |{1}\rangle_{\textrm{c}} \langle{1} |\otimes \hat{U}^{\dagger} \hat{M}_k^{\dagger} \hat{U} \rho_{\textrm{tb}}^{\textrm{in}} \hat{U}^{\dagger} \hat{M}_k \hat{U}\right]. \nonumber
                \end{align}
            \end{widetext}
            Using Eq.~(\ref{eq: Weingarten}), we have
            \begin{equation} \label{eq: ico_raw}
                \rho^{\mathrm{out}}_{\textrm{ctb}} =  \hat{\mathbb{I}}_c \otimes [p \rho_{\textrm{tb}}^{\textrm{in}} + (1-p) \hat{\mathbb{I}}_{\textrm{tb}}/D_{\textrm{tb}}]/2
                + \hat{\sigma}_{c}^x \otimes \tilde{\rho}_{\textrm{tb}}/2,
            \end{equation}
            where $p = ({\sum_k |\mathrm{Tr} \hat{M}_k|^2 - 1})/({D_{\textrm{tb}}^2 -1})$, and the density matrix in the second term is
            \begin{align} \label{eq: ico_coherent}
                    \tilde{\rho}_{\textrm{tb}}= &\left(p + \frac{1}{D_{\textrm{tb}}^2 -1}\right) \rho_{\textrm{tb}}^{\textrm{in}} + \frac{1}{D_{\textrm{tb}}^2 - 1} \sum_k \hat{M}_k^{\dagger} \rho_{\textrm{tb}}^{\textrm{in}} \hat{M}_k \nonumber\\
                     - &\frac{1}{D_{\textrm{tb}}(D_{\textrm{tb}}^2 - 1)} \sum_k [\mathrm{Tr}(\hat{M}_k) \rho_{\textrm{tb}}^{\textrm{in}}\hat{M}_k^{\dagger} + \mathrm{Tr}(\hat{M}_k^{\dagger}) \hat{M}_k \rho_{\textrm{tb}}^{\textrm{in}}].
            \end{align}

            Here, we assume the large dimension limit  $D_{\textrm{tb}} \rightarrow \infty$.
            The output state is asymptotically obtained as (see Appendix~\ref{app: asymptotic})
            \begin{equation} \label{eq: output}
                \rho^{\mathrm{out}}_{\textrm{ctb}}  \rightarrow  \hat{\mathbb{I}}_\textrm{c} \otimes [p \rho^{\textrm{in}}_{\textrm{tb}}  + (1-p) \hat{\mathbb{I}}_{\textrm{tb}}/D_{\textrm{tb}}] /2
                    + \hat{\sigma}_\textrm{c}^x \otimes p\rho^{\textrm{in}}_{\textrm{tb}} /2. 
            \end{equation}
            By tracing out the bath, we obtain that
                 \begin{equation}
                \rho^{\mathrm{out}}_{\textrm{ct}}  \rightarrow  \hat{\mathbb{I}}_\textrm{c} \otimes [p \rho^{\textrm{in}}_{\textrm{t}}  + (1-p) \hat{\mathbb{I}}_{\textrm{t}}/D_{\textrm{t}}] /2
                    + \hat{\sigma}_\textrm{c}^x \otimes p\rho^{\textrm{in}}_{\textrm{t}} /2. 
            \end{equation}

            Since the reduced density matrix of $q_\textrm{c}$ can be written as
            $\rho^{\mathrm{out}}_{\textrm{c}}=(\hat{\mathbb{I}}_\textrm{c}+p\hat{\sigma}_\textrm{c}^x)/2$, by measuring the expectation $\langle\hat{\sigma}_\textrm{c}^x\rangle=p$ on the control qubit, we can
            obtain exactly the recovery rate $p$, which resolve the first problem.
            Then, for the measurement in the basis of $|\pm\rangle_\textrm{c}$,
            the post-selected states are
            \begin{equation} \label{eq: ico}
                \rho^{+}_{\textrm{t}} =
                    \frac{2p}{1 + p} \rho^{\textrm{in}}_{\textrm{t}} + \frac{1-p}{1+p} \frac{\hat{\mathbb{I}}_{\textrm{t}}}{D_{\textrm{t}}},
            \end{equation}
            and $\rho^{-}_{\textrm{t}}={\hat{\mathbb{I}}_{\textrm{t}}}/{D_{\textrm{t}}}$, when the outcomes are $\pm$ with probabilities
            $p_\pm=(1\pm p)/2$, respectively.
            Thus, by post-selecting the state $\rho^{+}$ with respect to the outcome $+$,
            we can distill the initial state in the mixture from a recovery rate $p$ to
            \begin{align}
              p_{\textrm{ICO}}=2p/(1+p)\geq p,
            \end{align}
            with a successful probability $p_{+} = ({1 + p})/{2}$.

            Therefore, this scheme, by performing the scrambling procedure in ICO with a control qubit, can record the damage and improve the fidelity of recovered state.
            Moreover, it does not require any information of the initial state of the bath qubits and  even works without bath qubits.

        \subsection{Measurement-direction-dependent (MDD) Scheme of Scrambling Recovery}
\begin{figure}[t]
                \centering
                \includegraphics[width=0.45\textwidth]{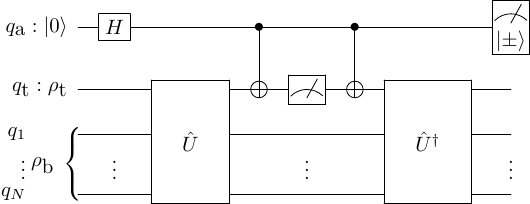}
                \caption{Quantum circuit for the measurement-direction-dependent (MDD) scheme for scrambling recovery of information.}
                \label{fig: direction}
            \end{figure}

            Then, we discuss a special and considerable case, in which the damage by Bob on the
            target qubit are assumed to be measurements with operators $\{\hat{\Pi}_s\}$ with
            $\sum_s\hat{\Pi}_s^\dag\hat{\Pi}_s=\hat{\mathbb{I}}$ and outcomes $s = \pm 1$.
            For this case, we can propose a measurement-direction-dependent (MDD) scheme that could completely recover the initial
            state of the target system for some measurement direction.
            The circuit of the MDD scheme is shown in Fig.~\ref{fig: direction},
            where an accilary qubit, $q_{\textrm{a}}$, is used to detect the direction of the measurement.
            Here, $\hat{U}$ is the random unitary operator, and the state after the measurement with a outcome $s=$ would be expressed as $\rho_{\textrm{atb}}^{(s)} = \hat{\Pi}_s^{\dagger} \rho_{\textrm{atb}} \hat{\Pi}_s/p_s$, for the outcome  with a probability $p_s=\textrm{Tr}[\hat{\Pi}_s^{\dagger} \rho_{\textrm{atb}} \hat{\Pi}_s]$, where $\bm{r} = (r_x, r_y,r_z)$ is the direction of the projective measurement.

            Then, the output state is
            \begin{align}
                   \rho^{(s)}_{\textrm{atb}} = & \Lambda_{\mathrm{twirl}} (\rho^{\mathrm{in}}_{\textrm{atb}} ) =  \frac{1}{2}|0\rangle _{\textrm{a}}\langle0|\otimes \rho^{(s)}_{\textrm{tb1}}
                    + \frac{1}{2}|0\rangle_{\textrm{a}} \langle{1}|\otimes \rho^{(s)}_{\textrm{tb2}}\nonumber\\
                    & + \frac{1}{2}|{1}\rangle_{\textrm{a}}\langle{0} |\otimes \rho^{(s)}_{\textrm{tb3}}
                    + \frac{1}{2}|{1}\rangle _{\textrm{a}}\langle1|\otimes \rho^{(s)}_{\textrm{tb4}},
\label{eq: c_twirling}
            \end{align}
            where
            \begin{subequations}
            \begin{align}
                   &  \rho_{\textrm{tb1}}^{(s)} \propto \int \!\!d\hat{U}\; \hat{U}^{\dagger} \hat{\Pi}_s^{\dagger} \hat{U} \rho^{\textrm{in}}_{\textrm{tb}} \hat{U}^{\dagger} \hat{\Pi}_s \hat{U},\\
                    & \rho_{\textrm{tb2}}^{(s)}\propto \int \!\!d\hat{U}\; \hat{U}^{\dagger} \hat{\Pi}_s^{\dagger} \hat{U} \rho^{\textrm{in}}_{\textrm{tb}} \hat{U}^{\dagger} \hat{\sigma}_{\textrm{t}}^x \hat{U}\hat{\Pi}_s \hat{\sigma}_{\textrm{t}}^x \hat{U}, \\
                    & \rho_{\textrm{tb3}}^{(s)} \propto \int \!\!d\hat{U}\; \hat{U}^{\dagger} \hat{\sigma}_{\textrm{t}}^x \hat{\Pi}_s^{\dagger} \hat{\sigma}_{\textrm{t}}^x \hat{U} \rho^{\textrm{in}}_{\textrm{tb}}  \hat{U}^{\dagger} \hat{\Pi}_s \hat{U}, \\
                    & \rho_{\textrm{tb4}}^{(s)}\propto \int \!\!d\hat{U}\; \hat{U}^{\dagger} \hat{\sigma}_{\textrm{t}}^x \hat{\Pi}_s^{\dagger} \hat{\sigma}_{\textrm{t}}^x \hat{U} \rho^{\textrm{in}}_{\textrm{tb}} \hat{U}^{\dagger} \hat{\sigma}_{\textrm{t}}^x \hat{\Pi}_s \hat{\sigma}_{1}^x \hat{U} .
            \end{align}
            \end{subequations}
            Applying Eq.~(\ref{eq: Weingarten}) (see Appendix~\ref{app: direction} for more details), we obtain that
            \begin{subequations}
                \begin{align}
                    \rho_{\textrm{tb1}}^{(s)} = \rho_{\textrm{tb4}}^{(s)} = \frac{1}{2}[p \rho^{\textrm{in}}_{\textrm{tb}} + (1 - p) {\hat{\mathbb{I}}_{\textrm{tb}}}/{D_{\textrm{tb}}}], \\
                    \rho_{\textrm{tb2}}^{(s)} = \rho_{\textrm{tb3}}^{(s)} = \frac{1}{2} [{p}' \rho^{\textrm{in}}_{\textrm{tb}} + (r_x^2 - {p}') {\hat{\mathbb{I}}_{\textrm{tb}}}/{D_{\textrm{tb}}}],
                \end{align}
            \end{subequations}
            where
            \begin{align}
                {p}' = \frac{D_{\textrm{tb}}^2/2 - r_x^2}{D_{\textrm{tb}}^2 - 1} \rightarrow p \rightarrow \frac{1}{2},~\textrm{for}~D_{\textrm{tb}} \rightarrow \infty.\nonumber
            \end{align}
            In the large dimension limit $D_{\textrm{tb}} \rightarrow \infty$, the output state with respect to the outcome $s$
            can be written as
            \begin{align}
                \rho^{(s)}_{\textrm{atb}} \rightarrow & \hat{\mathbb{I}}_{\textrm{a}} \otimes [p \rho^{\textrm{in}}_{\textrm{tb}} + (1 - p) {\hat{\mathbb{I}}_{\textrm{tb}}}/{D_{\textrm{tb}}}]/2\nonumber \\
                    & + \hat{\sigma}_{\textrm{a}}^x \otimes [p \rho^{\textrm{in}}_{\textrm{tb}} + \left(r_x^2 - p\right) {\hat{\mathbb{I}}_{\textrm{tb}}}/{D_{\textrm{tb}}}]/2 ,
            \end{align}
            and the corresponding probability $p_{s} = \frac{1}{2}$.
            Note that the output state is independent of the outcome $s$ of the projective measurement, which hereafter is omitted.

            By measuring the auxiliary qubit $q_{\textrm{a}}$ in the $\ket{\pm}_{\textrm{a}}$ basis, we could obtain
            \begin{equation}
                \rho^{+}_{\textrm{tb}} = \frac{2p}{1 + r_x^2} \rho^{\textrm{in}}_{\textrm{tb}} + \left(1 - \frac{2p}{1 + r_x^2}\right) \frac{\hat{\mathbb{I}}_{\textrm{tb}}}{D_{\textrm{tb}}},
            \end{equation}
            with a successful conditional probability $p_{+|s} = ({1+r_x^2})/{2}$, and $\rho^{-}_{\textrm{tb}} = \hat{\mathbb{I}}_{\textrm{tb}}/D_{\textrm{tb}}$ with a failing conditional probability $p_{-|s} = ({1 - r_x^2})/{2}$.
            Thus, the squred $x$-component of the measurement $r_x^2$ can be estimated by measuring the auxiliary qubit in the $\ket{\pm}_{\textrm{a}}$ basis. Furthermore, by  post-selecting the state $\rho^{+}_{\textrm{tb}}$ with respect to the outcome $+$, the recovery rate is improved to
            \begin{align}
              p_{\textrm{MDD}} = {2p}/({1 + r_x^2}) \geq p,
            \end{align}
            with $0 \leq r_x^2 \leq 1$, and the output state of the target system is
            \begin{equation}
                \rho^{+}_{\textrm{t}} = \frac{2p}{1 + r_x^2} \rho^{\textrm{in}}_{\textrm{t}} + \left(1 - \frac{2p}{1 + r_x^2}\right) \frac{\hat{\mathbb{I}}_{\textrm{t}}}{D_{\textrm{t}}}
            \end{equation}
            Surprisingly, if $r_x = 0$, i.e., the direction of the measurement is in the $y$-$z$ plane, with $p = {1}/{2}$, we have $p_{\textrm{MDD}} = 1$, which indicates a complete recovery of the initial state.

            \begin{figure*}[t]
                \centering
                \includegraphics[width=0.6\textwidth]{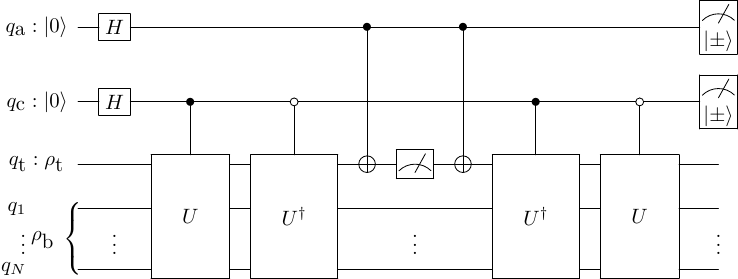}
                \caption{Quantum circuit for the combination of the ICO and MDD schemes.}
                \label{fig: double}
            \end{figure*}

            We also consider the case that the measurement is not in a fixed direction, e.g., the attacker perform measurement in random directions $r_x = \sin \theta$, with $\theta$ being uniformly distributed in the range $[0,2\pi]$. We can obtain the average $\overline{r_x^2}  = 1/2$, and the recovery rate is $\bar{p}_{\textrm{MDD}} = 2/3$, with averaged successful probability $\bar{p}_{+} = 3/4$, which is same as the ICO scheme.

            In summary, for a direction-fixed measurement, the MDD scheme can record the squared $x$-component of the measurement $r_x^2$, which allows us to optimize the scheme to completely recover initial quantum information of the target system.
            Even for the measurements in random directions, it behaves as good as the ICO scheme.

        \subsection{Combination of Indefinite Causal Order (ICO) and Measurement-Direction-Dependent (MDD) Schemes}

            The above two schemes are different:
            The ICO scheme requires a control qubit manipulating the  forwards and backwards evolutions, and the MDD scheme requires an ancillary qubit detecting the measurements on the target system.
            Thus, it is possible to combine  these two schemes, of which the quantum circuit is shown in Fig.~\ref{fig: double}.

            After the post-section with respect to the outcome $+_{\textrm{c}}+_{\textrm{a}}$ of the control qubit and the ancillary qubit,  the final state of the
            target system can calculated  as
            \begin{equation}
                \rho^{++}_{\textrm{t}} = \frac{4p}{1 + r_x^2 + 2p} \rho^{\textrm{in}}_{\textrm{t}} + \left(1 - \frac{4p}{1 + r_x^2 + 2p}\right) \frac{\hat{\mathbb{I}}_{\textrm{t}}}{D_{\textrm{t}}},
            \end{equation}
            with a successful probability
            $p_{++} = ({1+r_x^2})/{4} + {p}/{2}$.
            The rate $p_{\textrm{co}}$ of recovered information is
            \begin{equation}
                p_{\textrm{co}} = \frac{4p}{1 + r_x^2 + 2p} \geq \frac{2p}{1 + p},
            \end{equation}
            which is larger than either the ICO scheme or the MDD scheme.
            In addition, when $r_x = 0$ with $p ={1}/{2}$, we obtain that
            $p_{\textrm{co}} = 1$, which means the merit of total recovery is inherited successfully.
            Similarly, for the case that the measurement direction randomly chosen, the average recovery rate is $\bar{p}_{\textrm{co}} = {4}/{5}$, with an average successful probability
            $\bar{p}_{++} = {5}/{8}$.
            These results show that these the ICO and MDD schemes are well compatible with each other.

    \section{Iterative Scrambling Recovery Schemes} \label{sec: qudit}

        Next, we discuss iterative scrambling recovery schemes, which improve the fidelity between the initial state and the recovered state.
        For the scrambling recovery of the target system with $D_{\textrm{t}}$ dimension, the recover rate $p$ scales as
        $\sim D_{\textrm{t}}^{-l}$, where $l$ depends on the types of the perturbations on target system. It is possible to improve the recovery rate close to $1$ exponentially with an increasing number of iterations.
        However, the errors, occurring during the scrambling procedure, will ruin a complete recovery of initial information, which is similar to the EPP. Fortunately, the scrambling recovery schemes are compatible with QECC or EPP.

        \begin{figure}[b]
            \centering
            \includegraphics[width=0.4\textwidth]{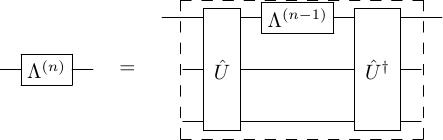}
            \caption{The circuit of iterative scrambling protocol.}
            \label{fig: iterate}
        \end{figure}

        \subsection{Iterative Scrambling Recovery Protocol}

        The random evolutions,  applied on the same target system and bath several times directly, are not effective
        to iterate the scrambling recovery protocol.
        The reason is that the evolution is random with Haar measure, which is invariant under the group multiplication of unitary evolution group $U(2^N)$. Thus, no matter how many layers of scrambling are applied, it is equivalent to a single layer of scrambling.
        The solution to this problem is to scramble the target system with different baths, the circuit is shown in Fig.~\ref{fig: iterate}.

        We consider a $n$-iterated twirling channel $\Lambda^{(n)}$ with the perturbation $ \Lambda$ acting on target system and set $\Lambda^{(0)} \equiv \Lambda$.
        For $n \geq 1$, the twirling channel $\Lambda^{(n-1)}$ can be regarded as a depolarizing channel with a depolarized rate $1 - p^{(n-1)}$, which can be written as
        \begin{align}
            &\Lambda^{(n-1)}(\rho_{\textrm{t}} )  = p^{(n-1)} \rho_{\textrm{t}} + (1-p^{(n-1}) \mathrm{Tr}(\rho_{\textrm{t}}) \hat{\mathbb{I}}_{\textrm{t}}/D_{\textrm{t}} \\
            &= \frac{(D_{\textrm{t}}^2 - 1)p^{(n-1)} + 1}{D_{\textrm{t}}^2} \hat{U}_0 \rho \hat{U}_{0}^{\dagger} + \frac{1-p^{(n-1)}}{D_{\textrm{t}}^2}  \sum_{i \neq 0} \hat{U}_{i} \rho_{\textrm{t}} \hat{U}_{i}^{\dagger} \nonumber,        \end{align}
        where $\{\hat{U}_{i}\}$ are the orthogonal unitary operator bases acting on the target system, which satisfy $\hat{U}_0 = \hat{\mathbb{I}}_{\textrm{t}}$, and $\mathrm{Tr}(\hat{U}_i^{\dagger} \hat{U}_j) = D_{\textrm{t}} \delta_{ij}$, for $i = 0, 1, \dots, D_{\textrm{t}}^2-1$.
        Then the depolarized rate of $\Lambda_{n}$ satisfies
        \begin{equation} \label{eq: iteration}
            1 - p^{(n)} = s (1 - p^{(n-1)}),
        \end{equation}
        where $s = \frac{1 - 1/D_{\textrm{t}}^2}{1 - 1/D_{\textrm{tb}}^2}$.
        When the dimension $D_{\textrm{t}}$ of the target system is clearly smaller than dimension $D_{\textrm{tb}}$ of total scrambled system, $s < 1$, the depolarized rate of depolarizing channel is suppressed after scrambling.
        Thus, we get
        \begin{equation}
            \rho^{(n)}_{\textrm{t}} = p^{(n)} \rho^{\textrm{in}}_{\textrm{t}} + (1 - p^{(n)})\hat{\mathbb{I}}_{\textrm{t}} /D_{\textrm{t}} \rightarrow \rho^{\textrm{in}}_{\textrm{t}} ,
        \end{equation}
        with the recovery rates being written as
        \begin{equation}
            p^{(n)} = 1-(1 - p) s^{n-1} \rightarrow1,
        \end{equation}
        where $p = ({\sum_k |\mathrm{Tr}\hat{M}_k|^2 - 1})/({D_{\textrm{tb}}^2 - 1})$ is determined by perturbation $\Lambda$.
        Thus, the output target state shows that the $n$-th iterated twirling channel can perform a complete recovery of quantum information of the target system.
        
        Imperfect scrambling will affect the performance of this scheme (for details, see Appendix~\ref{app: error}).
        It is shown that the fixed point under noises is
        \begin{equation}
            p^{(\infty)} = \frac{s x}{1 - (1 - s)x},
        \end{equation}
        where the noise is assumed to be a depolarizing error with a depolarizing rate $1-x$.
        When the depolarizing rate scales as $1 - x = \alpha D_{\textrm{t}}^{-2}$, the fixed point approaches to a constant $p^{(\infty)} \rightarrow {1}/{(1 + \alpha)}<1$, as $D_{\textrm{t}} \rightarrow \infty$.

        Note that this convergence is independent of the initial recovery rates $p$ decided by the form of the perturbation.
        The reason is that Eq.~(\ref{eq: iteration}) is a linear equation, which has only one attractive fixed point when $D_{\textrm{t}} < D_{\textrm{tb}}$. 
        Moreover, although the noise in scrambling will affect the fixed point of recovery rate, the noisy recursive equation is still a linear equation. There is no threshold where the fixed will vanish, which is different from the EPP or QECC.
        Therefore, this scheme is robust against the noises in scrambling.
        Here, we also remark that this scheme requires at least one bath qubit in each step of iteration, otherwise $D_{\textrm{t}} = D_{\textrm{tb}}$, and any value of $p$ is the fixed point.
        Thus, the number of baths will increase with iterations, which decides the performance of the recovery of  initial quantum information .

        \subsection{Iterative Indefinite Causal Order Scheme}

        \begin{figure*}[t]
            \centering
            \includegraphics[width=0.70\textwidth]{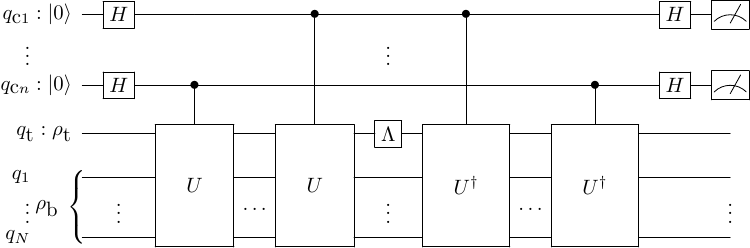}
            \caption{Quantum circuit for the asymptotic iterative ICO-scrambling scheme.}
            \label{fig: iterate_ico}
        \end{figure*}

        \begin{figure*}[t]
            \centering
            \includegraphics[width=0.82\textwidth]{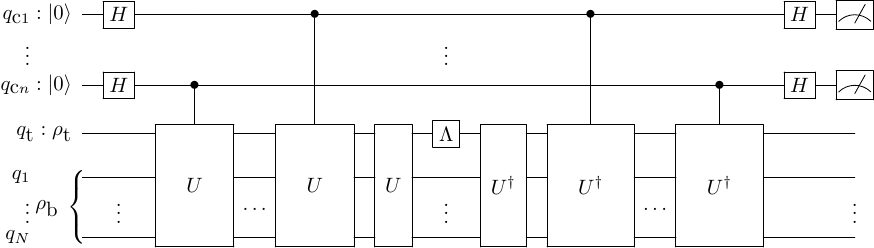}
            \caption{Quantum circuit for the exact iterative ICO-scrambling scheme.}
            \label{fig: iterate_ico_fin}
        \end{figure*}

        For the iterative ICO-scrambling scheme, we add $n$ control qubits, $q_{\textrm{c}_1},q_{\textrm{c}_2},q_{\textrm{c}_3},\cdots,q_{\textrm{c}_n}$, that manipulating the forward or backwards random evolution.
        The circuit of this scheme is shown in Fig.~\ref{fig: iterate_ico}.

        In the limit $D_{\textrm{tb}} \rightarrow \infty$, after post-selecting the state with respect to the outcome $+$ of the control qubit, the first ICO-twirling channel $\Lambda^{(1)}_{\textrm{ICO}}$ behaves as a depolarizing channel acting on the target system and the bath with a depolarized rate $(1 - p^{(1)}_{\textrm{ICO}})$, where
        \begin{equation}
            p^{(1)}_{\textrm{ICO}} = \frac{2 p}{1 + p},
        \end{equation}
        and $p$ is the recovery rate of the original perturbation $\Lambda^{(0)}_{\textrm{ICO}} \equiv \Lambda$ for the scrambling procedure.
        Then, we can assume that the $n-1$-th ICO-twirling channel $\Lambda^{(n-1)}_{\textrm{ICO}}$ also behaves as a depolarizing channel on target system and bath with a depolarized rate $(1 - p^{(n-1)}_{\textrm{ICO}})$ as
        \begin{align} \label{eq: depolarizing}
            & \Lambda^{(n-1)}_{\textrm{ICO}}(\rho_{\textrm{tb}} ) = p^{(n-1)}_{\textrm{ICO}} \rho_{\textrm{tb}} + (1-p^{(n-1)}_{\textrm{ICO}}) \mathrm{Tr}(\rho_{\textrm{tb}}) \hat{\mathbb{I}}_{\textrm{tb}}/D_{\textrm{tb}} \\
            & = \frac{(D_{\textrm{tb}}^2 - 1)p^{(n-1)}_{\textrm{ICO}} + 1}{D_{\textrm{tb}}^2} \hat{U}_0 \rho_{\textrm{tb}} \hat{U}_{0}^{\dagger} + \frac{1-p^{(n-1)}_{\textrm{ICO}}}{D_{\textrm{tb}}^2}  \sum_{i \neq 0} \hat{U}_{i} \rho_{\textrm{tb}} \hat{U}_{i}^{\dagger} \nonumber,
        \end{align}
        where $\{\hat{U}_{i}\}$ are orthogonal unitary operator bases acting on the target system and the bath,  with $\hat{U}_0 = \hat{\mathbb{I}}_{\textrm{tb}}$ and $\mathrm{Tr}(\hat{U}_i^{\dagger} \hat{U}_j) = D_{\textrm{tb}} \delta_{ij}$, for $i = 0, 1, \dots, D_{\textrm{tb}}^2-1$.
        Then, the recovery rate of the $n$-th iterative ICO-scrambling scheme can be obtained as
        \begin{equation}
            p^{(n)}_{\textrm{ICO}} = \frac{2^{n} p}{(2^{n}-1)p + 1} \rightarrow 1,
        \end{equation}
        from the recursive formula
        \begin{equation}
            p^{(n)}_{\textrm{ICO}} = \frac{2 p^{(n-1)}_{\textrm{ICO}}}{1 + p^{(n-1)}_{\textrm{ICO}}}.
        \end{equation}
        Therefore, the output state of the target system after the post-selection for the $n$-th iterative ICO scheme can be written as
        \begin{equation}
            \rho^{(n)}_{\textrm{t}} = \Lambda^{(n)}_{\textrm{ICO}}(\rho_{\textrm{t}}^{\textrm{in}}) = p^{(n)}_{\textrm{ICO}} \rho^{\textrm{in}}_{\textrm{t}} + (1 - p^{(n)}_{\textrm{ICO}})\hat{\mathbb{I}}_{\textrm{t}} /D_{\textrm{t}} \rightarrow \rho^{\textrm{in}}_{\textrm{t}},
        \end{equation}
        indicating a complete recovery of initial information.

        However, when the dimension of the bath is not infinite large, the side effect in the thermodynamic limit cannot be neglected. Thus, the iterative ICO-twirling channel constructed in Fig.~\ref{fig: iterate_ico} cannot be taken as a depolarizing channel.
        To overcome this problem, we first twirl the perturbation $\Lambda$ by scrambling it with the bath without using ICO, which is labelled as $\Lambda^{(0)}\equiv\Lambda_{\mathrm{twirl}}$ and can be expressed by a depolarization channel on the target system and the bath with a depolarizing rate $(1 - p)$, as shown in Eq.~(\ref{eq: twirling}). This exact iterative ICO-scrambling scheme is shown in Fig.~\ref{fig: iterate_ico_fin}.
        Then, we can obtain the $n-1$-th ICO-twirling channel as a depolarizing channel in Eq.~(\ref{eq: depolarizing}) for a finite size of the bath,  and the output state of the $n$-th ICO-twirling channel is given using Eq.~(\ref{eq: ico_raw}) and Eq.~(\ref{eq: ico_coherent}) as
        \begin{align}
            \rho^{(n)}_{\textrm{ctb}} =& \ \hat{\mathbb{I}}_c \otimes [p_{\textrm{eICO}}^{(n-1)} \rho_{\textrm{tb}}^{\textrm{in}} + (1-p_{\textrm{eICO}}^{(n-1)}) \hat{\mathbb{I}}_{\textrm{tb}}/D_{\textrm{tb}}]/2 \nonumber\\
            & + \hat{\sigma}_{c}^x \otimes \tilde{\rho}^{(n)}_{\textrm{tb}}/2,\label{eq: ico_iterative}\\
            \tilde{\rho}_{\textrm{tb}}^{(n)} =& \left[p_{\textrm{eICO}}^{(n-1)} + \frac{D_{\textrm{tb}}^2 - 2}{D_{\textrm{tb}}^2(D_{\textrm{tb}}^2 - 1)}(1 - p_{\textrm{eICO}}^{(n-1)})\right] \rho_{\textrm{tb}}^{\textrm{in}} \nonumber\\
            &+ \frac{1 - p^{(n-1)}}{D_{\textrm{tb}}^2 - 1} \frac{\hat{\mathbb{I}}}{D_{\textrm{tb}}}.\label{eq: ico_coherent_iterative}
        \end{align}
        By finally post-selecting the outcome of the control qubit $q_{\textrm{c}_n}$ as $+$ with a successful probability
        \begin{equation}
          p_+ = \frac{1 + p_{\textrm{eICO}}^{(n-1)}}{2} + \frac{1 - p_{\textrm{eICO}}^{(n-1)}}{D^2_{\textrm{tb}}},
        \end{equation}
        the output state can also be expressed as a density matrix through
        a depolarization channel as
        \begin{equation}
            \Lambda^{(n)}_{\textrm{eICO}}(\rho_{\textrm{t}} ) = p_{\textrm{eICO}}^{(n)} \rho_{\textrm{t}}^{\textrm{in}} + (1-p_{\textrm{eICO}}^{(n)}) \frac{\hat{\mathbb{I}}_{\textrm{t}}}{D_{\textrm{t}}},
        \end{equation}
        where
        \begin{equation}
            p_{\textrm{eICO}}^{(n)} = \frac{2 p_{\textrm{eICO}}^{(n-1)} + \frac{D_{\textrm{tb}}^2 - 2}{D_{\textrm{tb}}^2(D_{\textrm{tb}}^2 - 1)}(1 - p_{\textrm{eICO}}^{(n-1)})}{1 + p_{\textrm{eICO}}^{(n-1)} + \frac{2}{D_{\textrm{tb}}^2}(1 - p_{\textrm{eICO}}^{(n-1)})}.
        \end{equation}
        This recursive formula has an attractive fixed point at $p_{1}^{(\infty)} = 1$ and an unstable fixed point at $p_{2}^{(\infty)} = - {1}/({D_{\textrm{tb}}^2 - 1})$. Thus, the initial state can be recovered completely for the scrambling recovery scheme for any perturbation with a recovery rate $p > 0$.
        In the vicinity of $p_{1}^{(\infty)} = 1$,  the recursive equation up to linear order is written as
        \begin{equation}
            (p_{\textrm{eICO}}^{(n)} - 1) = \frac{1}{2}\left[1 + \frac{1}{(D_{\textrm{tb}}^2 - 1)}\right] (p_{\textrm{eICO}}^{(n-1)} - 1),
        \end{equation}
        where the coefficient
        \begin{equation}
            \frac{1}{2}\left[1 + \frac{1}{(D_{\textrm{tb}}^2 - 1)}\right] \leq \frac{2}{3},
        \end{equation}
        for $D_{\textrm{tb}} \geq 2$.
        While for iterative scheme without using ICO, the coefficient of linear order
        \begin{equation}
            \frac{1 - 1/D_{\textrm{t}}^2}{1 - 1/D_{\textrm{tb}}^2} \geq \frac{3}{4} \left(\frac{D_{\textrm{tb}}^2}{D_{\textrm{tb}}^2 - 1}\right) \geq \frac{3}{4},
        \end{equation}
        which means that the convergence of the iterative ICO-scrambling scheme to $1$ is faster than the standard iterative scrambling protocol for a weak perturbation $(1 - p_{\textrm{eICO}}^{(n)}) \ll 1$.

        The influence of the noise of the control qubits on this scheme is discussed in Appendix~\ref{app: error}.
        It is shown that the fixed points of the recovery rate are limited by the error rates on the control qubits.
        In addition, since the fixed point equation is of second order, there is a threshold for recovery, where the fixed points will vanish.
        This behavior is similar to the famous threshold of QECCs~\cite{Dur2007aa}.
        In comparison with iterative scrambling scheme without ICO, the iterative ICO-scrambling scheme can repeatedly use the bath, and is more fast in convergence than the iterative scrambling scheme without ICO, but is not robust against the noise on control qubits.

    \section{Quantum simulation of Scrambling Recovery Schemes with Virtual Baths with On-Cloud Superconducting Processor} \label{sec: simulation}

        \begin{figure}[b]
            \centering
            \includegraphics[width=8cm]{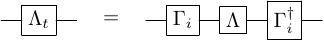}
            \caption{Quantum circuit for the $2$-design twirling channel.}
            \label{fig: design}
        \end{figure}

        \begin{figure}[t]
            \centering
            \subfigure[
            ]{
                \includegraphics[width=8cm]{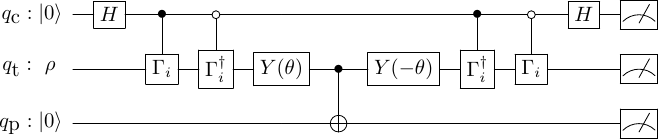}
                \label{subfig: design_ico}
            }
            \subfigure[
            ]{
                \includegraphics[width=8cm]{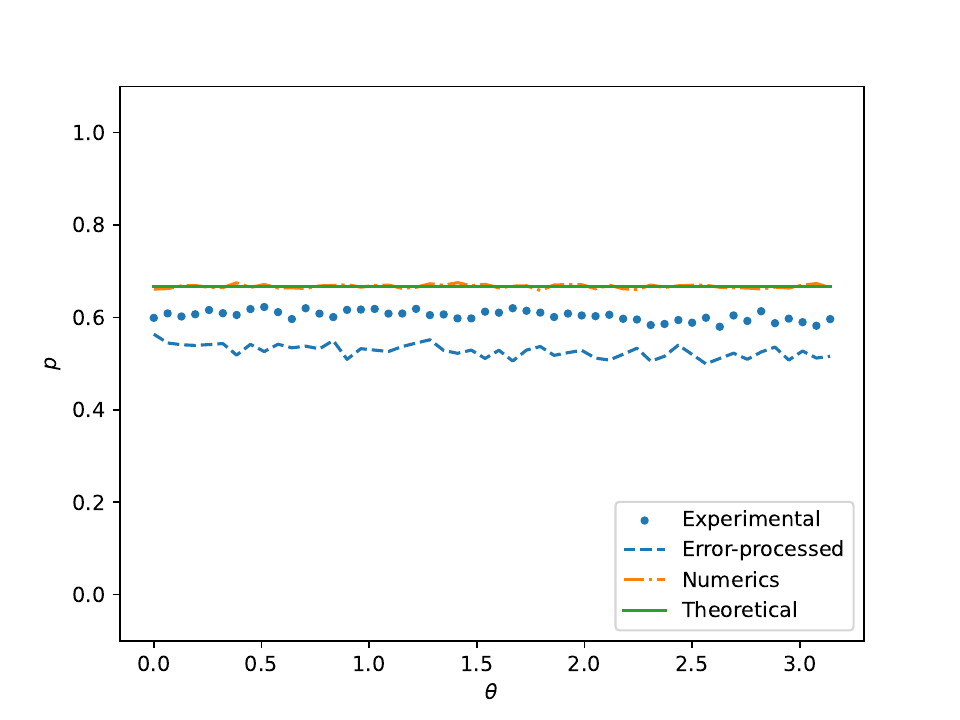}
                \label{subfig: trace}
            }
            \subfigure[%
            ]{
                \includegraphics[width=8cm]{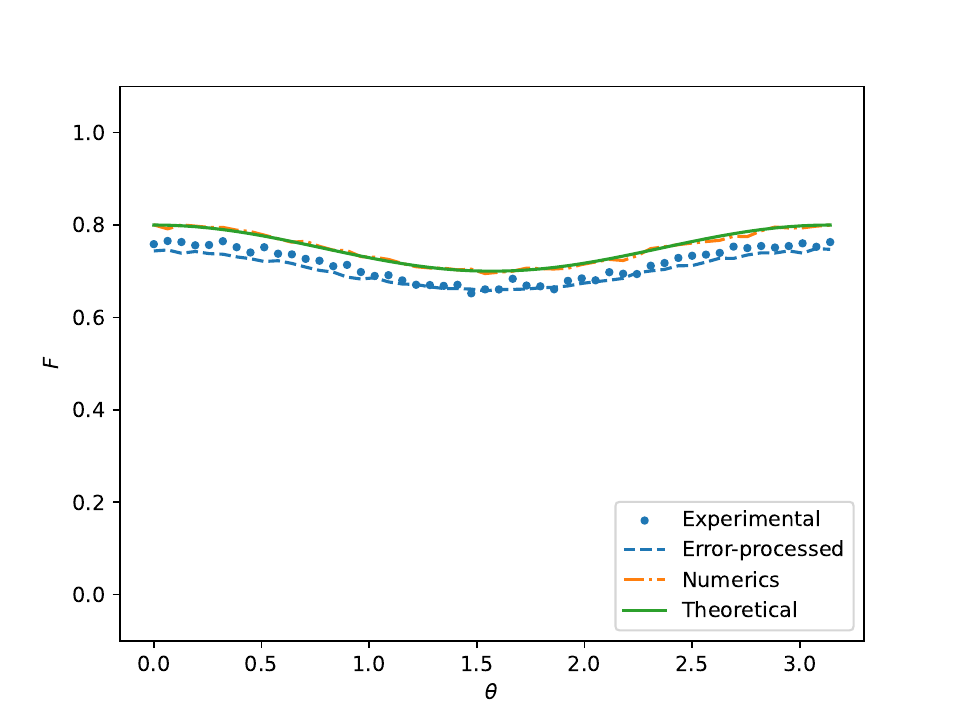}
                \label{subfig: fidelity_ico}
            }
            \caption{
                \ref{sub@subfig: design_ico} Quantum circuit for the ICO-scrambling recovery scheme via using the $2$-design.
                The probe qubit $q_{\textrm{p}}$ is used to perform the perturbative measurement on the target qubit $q_{\textrm{t}}$.
                The quantum circuit is performed in each element of Clifford group with $1,000$ shots, and the measurement angle $\theta$ is chosen in the range $[0, \pi]$ as $50$ equal spacing points.
                \ref{sub@subfig: trace} The recovery rate $p = \langle \hat{\sigma}_{\textrm{c}}^x \rangle$ versus the measurement angle $\theta$.
                \ref{sub@subfig: fidelity_ico} Fidelity of the recovered state  with the initial state $F$ versus $\theta$.
            }
            \label{fig: exp_ico}
        \end{figure}

        \begin{figure}[t]
            \centering
            \subfigure[
            ]{
                \includegraphics[width=8cm]{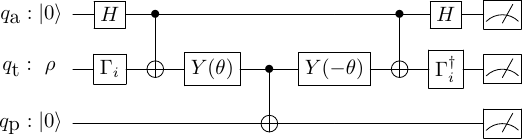}
                \label{subfig: design_direction}
                }
            \subfigure[
            ]{
                \includegraphics[width=8cm,clip=True]{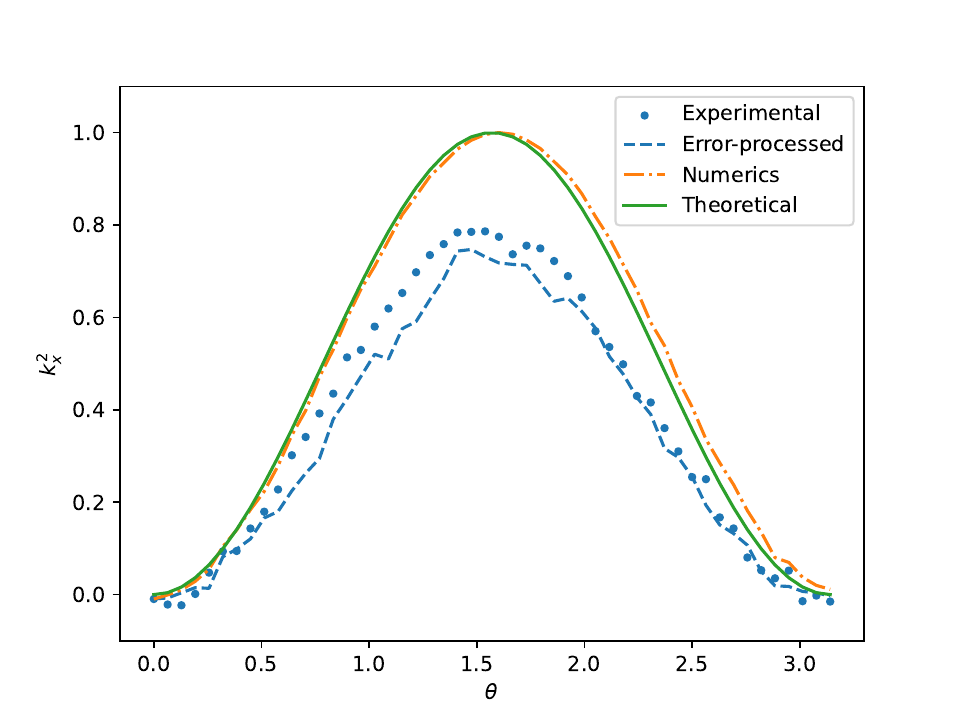}
                \label{subfig: direction}
            }
            \subfigure[
            ]{
                \includegraphics[width=8cm,clip=True]{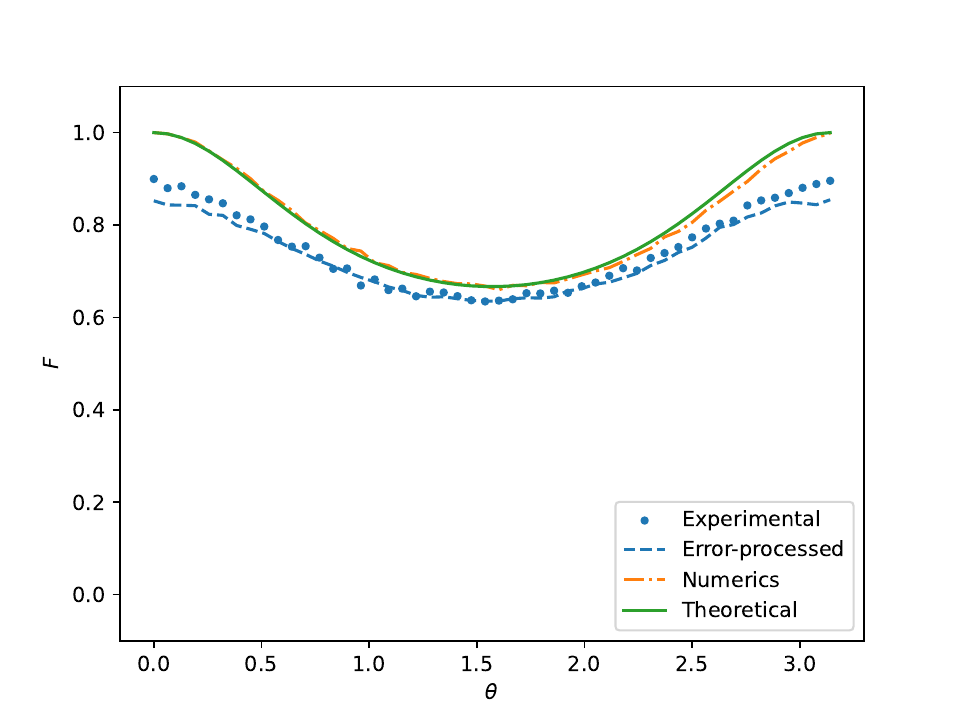}
                \label{subfig: fidelity}
            }
            \caption{
                \ref{sub@subfig: design_direction} Quantum circuit for the MDD-scrambling recovery scheme with the $2$-design.
                The probe qubit $q_{\textrm{p}}$ is used to perform the perturbative measurement on the target qubit $q_{\textrm{t}}$.
                The quantum circuit is performed in each element of Clifford group with $1,000$ shots, and the measurement angle $\theta$ is chosen in the range $[0, \pi]$ as $50$ equal spacing points.
                \ref{sub@subfig: direction} The $x$-component of the squared projective measurement $r_x^2 = \langle \hat{\sigma}_{\textrm{a}}^x \rangle$ versus the measurement angle $\theta$.
                \ref{sub@subfig: fidelity} Fidelity of the recovered state with the initial state versus $\theta$.
            }
            \label{fig: exp_anisotropic}
        \end{figure}

        \begin{figure*}[t]
            \centering
            \subfigure[
            ]{
                \includegraphics[width=12cm]{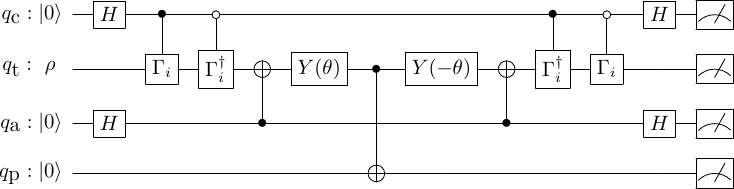}
                \label{subfig: design_direction_ico}
                }
            \subfigure[
            ]{
                \includegraphics[width=8cm,clip=True]{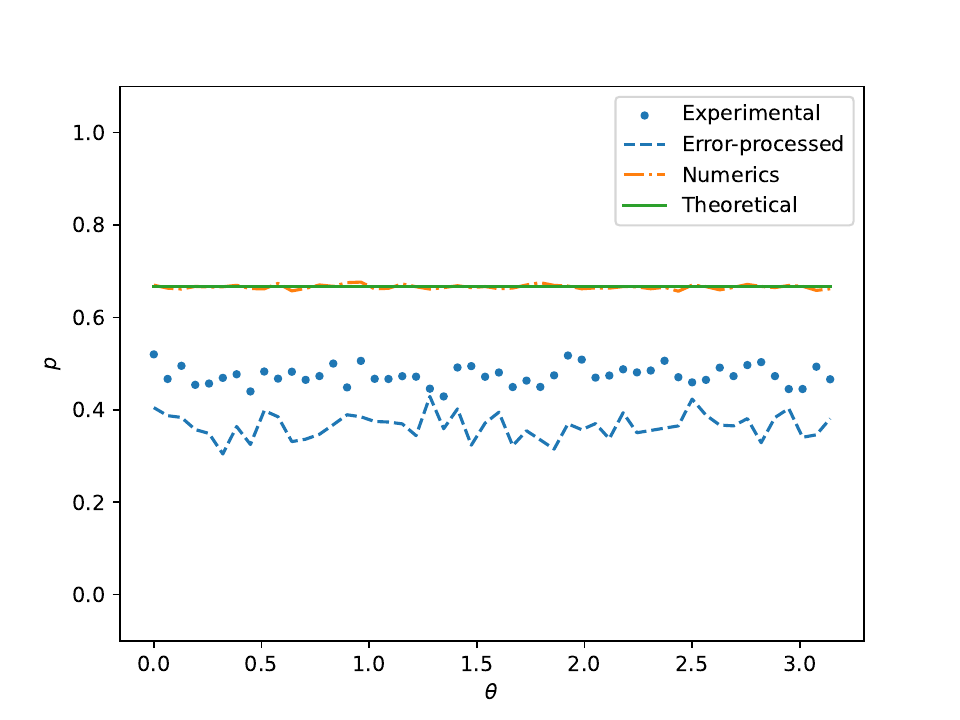}
                \label{subfig: trace_ico_d}
            }
            \subfigure[
            ]{
                \includegraphics[width=8cm,clip=True]{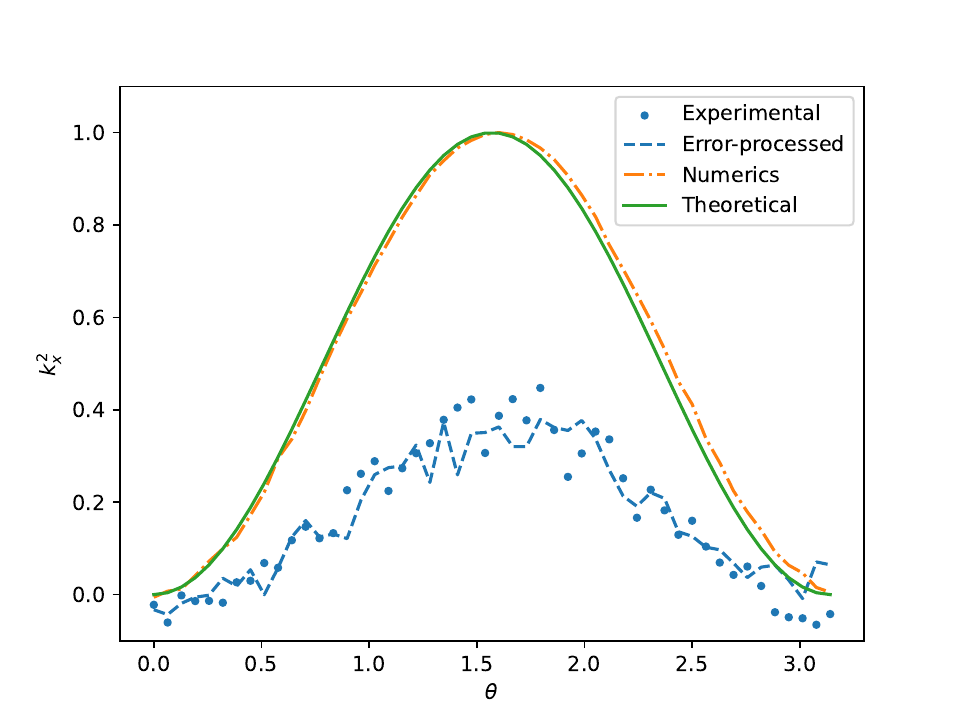}
                \label{subfig: r_x^2_ico_d}
            }
            \subfigure[
            ]{
                \includegraphics[width=8cm,clip=True]{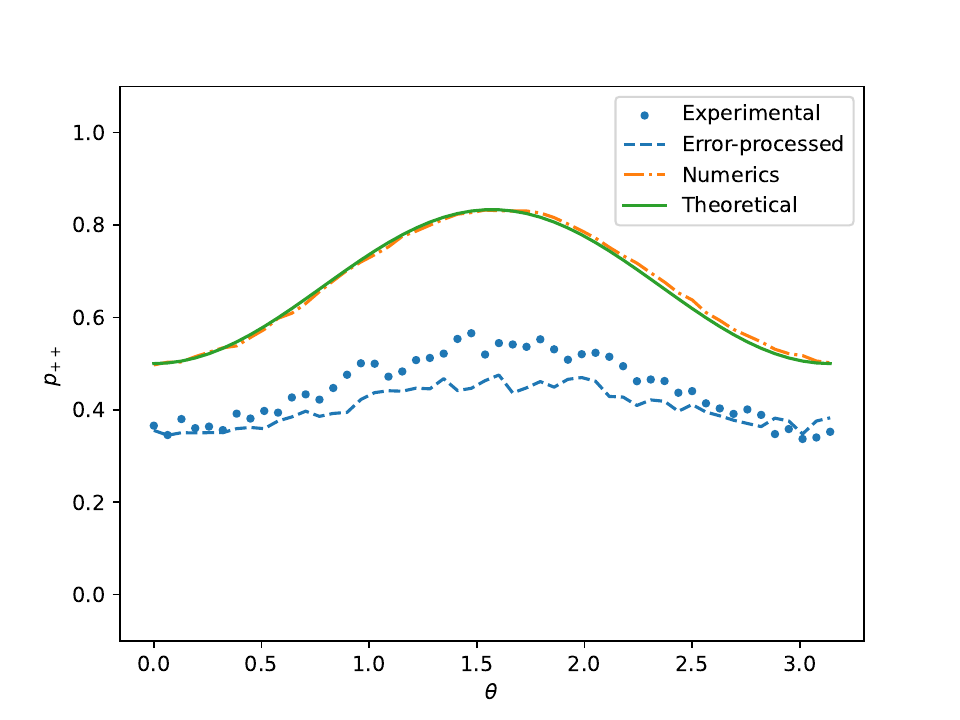}
                \label{subfig: q_ico_d}
            }
            \subfigure[
            ]{
                \includegraphics[width=8cm,clip=True]{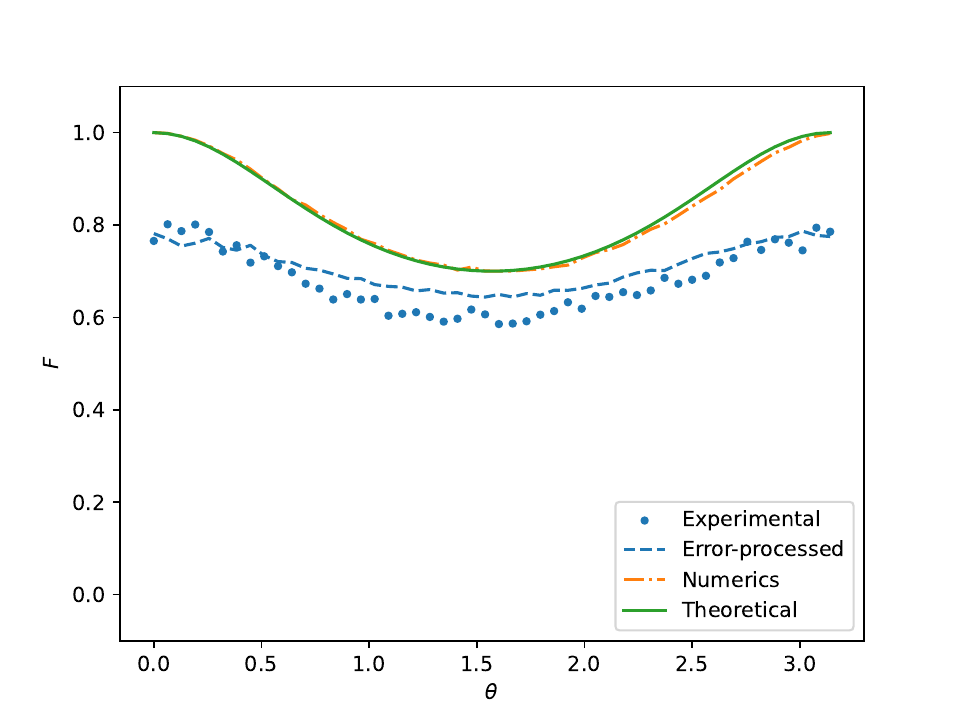}
                \label{subfig: fidelity_ico_d}
            }
            \caption{
                \ref{sub@subfig: design_ico} Quantum circuit for the combination of the ICO- and MDD-scrambling schemes with th $2$-design.
                The probe qubit $q_{\textrm{p}}$ is used to perform the perturbative measurement on the target qubit $q_{\textrm{t}}$.
                The quantum  circuit is performed in each element of the Clifford group with $1,000$ shots, and the measurement angle $\theta$ is chosen in the range $[0, \pi]$ as $50$ equally spacing points.
                \ref{sub@subfig: trace_ico_d} Recovery rate $p =\langle \hat{\sigma}_{\textrm{c}}^x \rangle$ versus the measurement angle $\theta$.
                \ref{sub@subfig: r_x^2_ico_d} The curve of direction $r_x^2 = \langle \hat{\sigma}_{\textrm{a}}^x \rangle$ versus $\theta$.
                \ref{sub@subfig: q_ico_d} Sucessful probability for the post-selection $p_{++}$ versus $\theta$.
                \ref{sub@subfig: fidelity_ico} Fidelity of the recovered state with the initial state versus $\theta$.
            }
            \label{fig: exp_ico_d}
        \end{figure*}

        The realization of a random unitary evolution with Haar measure is of great importance in information scrambling.
        However, to realize this randomness requires many layers of quantum gates, which is limited by the coherent time in the noisy intermediate-scale quantum (NISQ) era. Thus, a direct experimental realization of the randomness is not easy in state-of-art experimental technique.

        One way to preserve the property of randomness and to reduce the complexity of the quantum circuit simultaneously is to use the unitary $t$-design \cite{gross2007evenly}.
        For the twirling channel, we need at least to apply the $2$-design, and fortunately, the Clifford group, widely used in error bench-marking of quantum gates, satisfies this requirement (see Appendix~\ref{app: design} for more details).
        Here, we design the circuit of the twirling channel as shown in Fig.~\ref{fig: design},
        where $\Gamma_i$ is an element of the Clifford group, and $\sum_i$ represents the average on Clifford group.

        With this circuit, we can simulate our improved recovery schemes on a NISQ quantum processor.
        Since the Clifford group of $2$ qubits has $11,520$ elements~\cite{PhysRevA.87.030301} and is harder to realize than single qubit gates, only the schemes, in specific the ICO-  and  MDD-scrambling schemes, with virtual baths are simulated.
        The quantum circuits are performed on the \emph{Quafu} on-cloud quantum computers \cite{chen2022scq}, of which the experimentally results are compared to both  theoretical predictions and  numerical calculations.

        \subsection{Quantum Simulation of the ICO-scrambling Scheme}

        The quantum circuit is shown in Fig.~\ref{subfig: design_ico}, where $q_{\textrm{c}}$ is a control qubit, $q_{\textrm{t}}$ is the target qubit, and $q_{\textrm{p}}$ is the probe qubit used to perform the perturbative measurement.
        Here, $Y(\theta)$ denotes the rotation operators with an angle $\theta$ along $y$-axis, which determines the direction of the measurement.
        The initial state of the target system is prepared as $\ket{0_z}$.
        When averaging over the Clifford group, the measurement outcomes of the quantum circuit with the initial state $\ket{0_z}$ is equivalent to the outcomes with the initial states congruent to $\ket{0_z}$ up to the Clifford group, e.g., $\ket{0_x}$ and $\ket{0_y}$.

        The results are shown in Fig.~\ref{fig: exp_ico}.
        Figure~\ref{subfig: trace} shows that the expectation $\langle \hat{\sigma}_{\textrm{c}}^{x} \rangle$ of the control qubit $q_{\textrm{c}}$, representing the coherent part of the reduced density matrix of $q_{\textrm{c}}$, is invariant versus the direction $\theta$.
        In Fig.~\ref{subfig: fidelity_ico}, we show the fidelity versus the measurement direction $\theta$.
        Note that the fidelity varies for different measurement directions, which is different from the theoretical prediction in the previous discussions.
        This is because that the thermodynamic limit $D_{\textrm{tb}} \rightarrow \infty$ is not saturated, and the side effect that we have neglected accounts for the dependency of the fidelity on the measurement direction.
        The theoretical prediction, considering the side effect (see Appendix~\ref{app: simulation}) is plotted as  the solid curve in Fig.~\ref{fig: exp_ico}, which coincides with the numerical calculation on classical computer.

        The experimental results fit well with our theoretical prediction and numerical calculation even with some errors.
        We calibrate the noises with quantities $\braket{\hat{\sigma}_{\textrm{c}}^{x}}_{\pm}$ and $F^{\pm}$ measured from auxiliary circuits, where the perturbations $\Lambda$ are not applied.
        We do not process the data from the experiments to mitigate the noise.
        We take the calibrated noises into the consideration when performing the theoretical predictions (see Appendix~\ref{app: simulation} for more details).
        The theoretical results with noises are shown using the blue dashed curve in Fig.~\ref{fig: exp_ico}.

    \subsection{Quantum simulation of the MDD-scrambling Scheme}

        The quantum circuit is shown in Fig.~\ref{subfig: design_direction},
        where $q_{\textrm{a}}$ is an auxiliary qubit, $q_{\textrm{t}}$ is the target qubit, and $q_{\textrm{p}}$ is the probe qubit.
        The initial state of the target system is also prepared as $\ket{0_z}$.
        The direction of measurement, detected by the auxiliary qubit $q_{\textrm{a}}$, is shown in Fig.~\ref{subfig: direction}, and the fidelity of the recovered state of $q_{\textrm{t}}$ with initial state is shown in Fig.~\ref{subfig: fidelity}.
        The numerical calculation coincides with the theoretical prediction, and the experimental results also verifies the dependency of recovery fidelity on the measurement direction, even when the error is considered.

        We also show that the error depends on the measurement direction,  since only the $Y (\pi)$ gate of the on-cloud quantum processor has been carefully calibrated, while the $Y(\theta)$ gate with an arbitrary rotation angle $\theta$ has not.
        The theoretical results with noises are shown as the blue dashed curve in Fig.~\ref{fig: exp_anisotropic} (see Appendix~\ref{app: simulation} for more details).

    \subsection{Quantum Simulation of the Combination of the ICO- and MDD-scrambling Schemes}

        The quantum circuit for the combination of ICO- and MDD- scrambling schemes is shown in Fig.~\ref{subfig: design_direction_ico}, where $q_{\textrm{c}}$ is the control qubit, $q_{\textrm{t}}$ is the target qubit, $q_{\textrm{a}}$ is the auxiliary qubit, and $q_{\textrm{p}}$ is the probe qubit.
        The results are shown in Fig.~\ref{fig: exp_ico_d}.  Figure~\ref{subfig: q_ico_d} plotts the probability $p_{++}$ of the basis $\ket{+}_{\textrm{c}} \otimes \ket{+}_{\textrm{t}}$ on $q_{\textrm{c}}$ and $q_{\textrm{t}}$ for the post-selection.
        We see that the error between experimental and theoretical results is larger than the combination of the errors of two experiments above.
        Note that except for the direction dependent error appearing in the quantum simulation of the MDD-scrambling scheme, there is a large error that is independent of the direction.
        However, this error does not exist in the quantum simulation of the ICO-scrambling scheme, which is much smaller.

        The possible origin of this error is from the CNOT gate on $q_{\textrm{t}}$ and $q_{\textrm{p}}$.
        Since our quantum processor is a qubit chain, $q_{\textrm{p}}$ does not connect $q_{\textrm{t}}$ directly.
        Thus, to perform this CNOT gate, there require two SWAP gates between $q_{\textrm{a}},q_{\textrm{p}}$, which lead to a large error than the results of previous schemes.
        The theoretical results with noises are shown as the blue dashed line in Fig.~\ref{fig: exp_ico_d} (see Appendix~\ref{app: simulation} for more details).

    \section{Conclusion} \label{sec: conclusion}

        In this paper, we propose two improved  information recovery schemes via scrambling that overcome the problems encountered in the original recovery protocol proposed in  Ref.~\cite{PhysRevLett.125.040605}.

        The ICO-scrambling scheme asymptotically can record information of the damage and promote the fidelity of the recovered state by performing the post-selection.
        However, it has the side effect when the bath is not large enough.
        This drawback can be overcome by scrambling the target system without ICO before performing the ICO-scrambling scheme.
        The MDD-scrambling scheme is designed for the projective measurement perturbation on the qubit target system only.
        This scheme allows a complete recovery of the initial state, if the measurement direction is fixed in the $y$-$z$ plane. Otherwise, it allows for recording the $x$-component of the measurement.
        If the measurement direction is not fixed, the average recovery performance of the MDD-scrambling scheme is same as the ICO-scrambling scheme.

        In addition, these scrambling recovery schemes can also be iterated, to completely recover the initial state under arbitrary damage.
        We discuss the iteration of the original recovery protocol \cite{PhysRevLett.125.040605}, and the iterative ICO-scrambling scheme.
        To iterate the original scrambling recovery protocol would scramble the target system with different baths.
        Therefore, the iterative ICO-scrambling scheme could exhibit its advantage that it repeatedly uses a single bath.

        These scrambling recovery schemes use redundant qubits to reduce the effects of the damage on the target system, which in fact could be taken as a kind of passive QECs \cite{PhysRevA.55.900}.
        In comparison, the active QECs can improve the ability to defend the errors by detecting them and performing corrections based on their syndromes, and the types of errors are confined by the stabilizers \cite{PhysRevLett.78.405}.

        Finally, the ICO-scrambling scheme, the MDD-scrambling scheme, and their combination are simulated on the on-cloud quantum computer, \emph{Quafu}.
        The experimental results are compared with the theoretical predictions and numerical calculations, which fit well when error is processed.

        Our results provide methods to record the type of the damage and to distill the damaged quantum state, which allows to retrieve initial information against different types of the damage.
        We expect that our scheme will be useful in the both quantum information recovery from the damage and system's bench-marking.

    \begin{acknowledgments}
        This work was supported by National Natural Science Foundation of China (Grants Nos.92265207, T2121001, 11934018, 12122504), the Innovation Program for Quantum Science and Technology (Grant No. 2021ZD0301800), Beijing Natural Science Foundation (Grant No. Z200009), and Scientific Instrument Developing Project of Chinese Academy of Sciences (Grant No. YJKYYQ20200041).
        We also acknowlege the supported from the Synergetic Extreme Condition User Facility (SECUF) in Beijing, China.
    \end{acknowledgments}

    \section*{Data Availability}

    The data that support the findings of this study are openly available at \href{https://doi.org/10.6084/m9.figshare.25990429.v1}{https://doi.org/10.6084/m9.figshare.25990429.v1}.

    \begin{widetext}
    \appendix
    \section{Asymptotic Analysis of the recovered state} \label{app: asymptotic}
        To analyze the asymptotic behavior of $\tilde{\rho}_{\textrm{tb}}$, we employ the Hilbert-Schmidt inner product $(\hat{A}, \hat{B}) = \mathrm{Tr}(\hat{A}^{\dagger} \hat{B})$ and the induced norm $\Vert \hat{A} \Vert^2 = \mathrm{Tr}(\hat{A}^{\dagger} \hat{A})$.
        We have
        \begin{align}
        \Vert \tilde{\rho}_{\textrm{tb}} - p \rho_{\textrm{tb}}^{\textrm{in}} \Vert \leq &\left\Vert \frac{1}{D_{\textrm{tb}}^2 -1} \rho_{\textrm{tb}}^{\textrm{in}} \right\Vert + \left\Vert \frac{1}{D_{\textrm{tb}}^2 - 1} \sum_k \hat{M}_k^{\dagger} \rho_{\textrm{tb}}^{\textrm{in}} \hat{M}_k \right\Vert \\
        \end{align}
        Precisely, we have
        \begin{align}
        \left\Vert \frac{\rho_{\textrm{tb}}^{\textrm{in}}}{D_{\textrm{tb}}^2 -1} \right\Vert^2 = \frac{\mathrm{Tr}(\rho_{\textrm{tb}}^{\textrm{in}})^2}{D_{\textrm{tb}}^2 -1}  \leq \frac{1}{D_{\textrm{tb}}^2 -1} \rightarrow 0,
        \end{align}
        \begin{align}
            \left\Vert \frac{1}{D_{\textrm{tb}}^2 - 1} \sum_k \hat{M}_k^{\dagger} \rho_{\textrm{tb}}^{\textrm{in}} \hat{M}_k \right\Vert^2 &= \frac{1}{D_{\textrm{tb}}^2 - 1} \mathrm{Tr}\left[\left(\sum_k \hat{M}_k^{\dagger} \rho_{\textrm{tb}}^{\textrm{in}} \hat{M}_k\right)^2\right]\leq \frac{1}{D_{\textrm{tb}}^2 - 1} \rightarrow 0,
        \end{align}
        \begin{align}
                & \left\Vert \frac{1}{D_{\textrm{tb}}(D_{\textrm{tb}}^2 - 1)} \sum_k [\mathrm{Tr}(\hat{M}_k) \rho_{\textrm{tb}}^{\textrm{in}}\hat{M}_k^{\dagger} + \mathrm{Tr}(\hat{M}_k^{\dagger}) \hat{M}_k \rho_{\textrm{tb}}^{\textrm{in}}] \right\Vert\nonumber\\
                & = \frac{1}{D_{\textrm{tb}}(D_{\textrm{tb}}^2 - 1)} \sum_{k,l} \textrm{Re}[\mathrm{Tr}(\hat{M}_k) \mathrm{Tr}(\hat{M}_l) \mathrm{Tr}(\rho_{\textrm{tb}}^{\textrm{in}} \hat{M}_l^{\dagger} \rho_{\textrm{tb}}^{\textrm{in}} \hat{M}_k^{\dagger}) + \mathrm{Tr}(\hat{M}_k) \mathrm{Tr}(\hat{M}_l^{\dagger}) \mathrm{Tr}(\hat{M}_l \rho^2 \hat{M}_k^{\dagger})]\nonumber \\
                & \leq \frac{1}{D_{\textrm{tb}}(D_{\textrm{tb}}^2 - 1)} \sum_{k,l} |\mathrm{Tr}(\hat{M}_k) \mathrm{Tr}(\hat{M}_l)|\times \left[|\mathrm{Tr}(\rho_{\textrm{tb}}^{\textrm{in}} \hat{M}_l^{\dagger} \rho_{\textrm{tb}}^{\textrm{in}} \hat{M}_k^{\dagger})| + |\mathrm{Tr}(\hat{M}_l (\rho_{\textrm{tb}}^{\textrm{in}})^2 \hat{M}_k^{\dagger})|\right]\nonumber \\
                & \leq \frac{1}{D_{\textrm{tb}}(D_{\textrm{tb}}^2 - 1)} \left(\sum_k |\mathrm{Tr}(\hat{M}_k)|^2 \right) \left[\sqrt{\sum_{k,l} |\mathrm{Tr}(\rho_{\textrm{tb}}^{\textrm{in}} \hat{M}_l^{\dagger} \rho_{\textrm{tb}}^{\textrm{in}} \hat{M}_k^{\dagger})|^2} + \sqrt{\sum_{k,l} |\mathrm{Tr}(\hat{M}_l (\rho_{\textrm{tb}}^{\textrm{in}})^2 \hat{M}_k^{\dagger})|^2}\right].
            \end{align}
        Then, with the Cauchy-Schwarz inequality on Hilbert-Schmidt inner product, we have
            \begin{align}
            \sum_k |\mathrm{Tr}(\hat{M}_k)|^2
            &\leq \sum_k \mathrm{Tr}(\hat{\mathbb{I}}^2) \mathrm{Tr}(\hat{M}_k^{\dagger} \hat{M}_k) = D_{\textrm{tb}}^2, \\
            \sum_{k,l} |\mathrm{Tr}(\rho_{\textrm{tb}}^{\textrm{in}} \hat{M}_l^{\dagger} \rho_{\textrm{tb}}^{\textrm{in}} \hat{M}_k^{\dagger})|^2
            &\leq \sum_k \mathrm{Tr}(\rho_{\textrm{tb}}^{\textrm{in}} \hat{M}_k^{\dagger} \hat{M}_k \rho_{\textrm{tb}}^{\textrm{in}}) \sum_k \mathrm{Tr}[\hat{M}_k (\rho_{\textrm{tb}}^{\textrm{in}})^2 \hat{M}_k^{\dagger}] = \mathrm{Tr}^2\left(\rho_{\textrm{tb}}^{\textrm{in}}\right)^2\leq 1, \\
            \sum_{k,l} |\mathrm{Tr}[\hat{M}_l (\rho_{\textrm{tb}}^{\textrm{in}})^2 \hat{M}_k^{\dagger}]|^2
            &\leq \left[\sum_k \mathrm{Tr}(\hat{M}_k (\rho_{\textrm{tb}}^{\textrm{in}})^2 \hat{M}_k^{\dagger})\right]^2 = \mathrm{Tr}^2(\rho_{\textrm{tb}}^{\textrm{in}})^2\leq 1,
        \end{align}
        and
        \begin{align}
          \left\Vert \frac{1}{D_{\textrm{tb}}(D_{\textrm{tb}}^2 - 1)} \sum_k [\mathrm{Tr}(\hat{M}_k) \rho_{\textrm{tb}}^{\textrm{in}}\hat{M}_k^{\dagger} + \mathrm{Tr}(\hat{M}_k^{\dagger}) \hat{M}_k \rho_{\textrm{tb}}^{\textrm{in}}] \right\Vert  &\leq \frac{4}{(D_{\textrm{tb}}^2 - 1)^2}  \rightarrow 0.
        \end{align}
        Thus, we conclude that
        \begin{equation}
            \lim_{D_{\textrm{tb}} \rightarrow \infty} \tilde{\rho}_{\textrm{tb}} = p \rho_{\textrm{tb}}^{\textrm{in}}.
        \end{equation}

    \section{Output state of the MDD Scheme} \label{app: direction}

        With Eq.~(\ref{eq: Weingarten}), we have
        \begin{align}
            \rho_{\textrm{tb1}}^{(s)} = \rho_{\textrm{tb4}}^{(s)} \propto & \left[\frac{|\mathrm{Tr} \hat{\Pi}_s|^2}{D_{\textrm{tb}}^2 - 1} - \frac{\mathrm{Tr} (\hat{\Pi}_s^{\dagger} \hat{\Pi}_s )}{D_{\textrm{tb}} (D_{\textrm{tb}}^2 - 1)}\right] \rho^{\textrm{in}}_{\textrm{tb}} + \left[\frac{D_{\textrm{tb}} \mathrm{Tr} (\hat{\Pi}_s^{\dagger} \hat{\Pi}_s)}{D_{\textrm{tb}}^2 - 1} - \frac{|\mathrm{Tr} \hat{\Pi}_s|^2}{D_{\textrm{tb}}^2 - 1}\right]\frac{\hat{\mathbb{I}}}{D_{\textrm{tb}}}. \\
            \rho_{\textrm{tb2}}^{(s)} = \rho_{\textrm{tb3}}^{(s)}
            \propto & \left[\frac{|\mathrm{Tr} \hat{\Pi}_s|^2}{D_{\textrm{tb}}^2 - 1} - \frac{\mathrm{Tr} (\hat{\Pi}_s^{\dagger} \hat{\sigma}_{\textrm{t}}^x \hat{\Pi}_s \hat{\sigma}_{\textrm{t}}^x)}{D_{\textrm{tb}} (D_{\textrm{tb}}^2 - 1)}\right] \rho^{\textrm{in}}_{\textrm{tb}} + \left[\frac{D_{\textrm{tb}} \mathrm{Tr} (\hat{\Pi}_s^{\dagger} \hat{\sigma}_{\textrm{t}}^x \hat{\Pi}_s \hat{\sigma}_{\textrm{t}}^x)}{D_{\textrm{tb}}^2 - 1} - \frac{|\mathrm{Tr} \hat{\Pi}_s|^2}{D_{\textrm{tb}}^2 - 1}\right]\frac{\hat{\mathbb{I}}}{D_{\textrm{tb}}}.
        \end{align}
        With $\hat{\Pi}_{s=\pm1} = ({\hat{ \mathbb{I}} +s\bm{n}\cdot \bm{\sigma}_{\textrm{t}}})/{2}$,
        \begin{align}
            \mathrm{Tr} \left(\hat{\Pi}_s^{\dagger} \hat{\Pi}_s \right) & = \mathrm{Tr} \hat{\Pi}_s = \frac{D_{\textrm{tb}}}{2}, \\
            \mathrm{Tr} (\hat{\Pi}_s^{\dagger} \hat{\sigma}_x \hat{\Pi}_s \hat{\sigma}_x)
            & = \mathrm{Tr} \left(\frac{{\hat{ \mathbb{I}} +s\bm{r}\cdot \bm{\sigma}_{\textrm{t}}}}{2} \frac{{\hat{ \mathbb{I}} +s\bm{r}'\cdot \bm{\sigma}_{\textrm{t}}}}{2} \right)
            = (1 + \bm{r} \cdot \bm{r}') \frac{D_{\textrm{tb}}}{4} =  \frac{r_x^2 D_{\textrm{tb}}}{2},
        \end{align}
        where $\bm{r}' = (r_x, - r_y, - r_z)$ is the vector obtained from $\bm{r}$ by rotating $\pi$ about $x$ direction.
        Thus, we have
        \begin{align}
            \rho_{\textrm{tb1}}^{(s)} &= \rho_{\textrm{tb4}}^{(s)} = \frac{1}{2}\left[p \rho^{\textrm{in}}_{\textrm{tb}} + (1 - p) \frac{\hat{\mathbb{I}}_{\textrm{tb}}}{D_{\textrm{tb}}}\right], \\
            \rho_{\textrm{tb2}}^{(s)} &= \rho_{\textrm{tb3}}^{(s)} = \frac{1}{2} \left[p' \rho^{\textrm{in}}_{\textrm{tb}} + \left(r_x^2 - p'\right) \frac{\hat{\mathbb{I}}_{\textrm{tb}}}{D_{\textrm{tb}}}\right],
        \end{align}
        where
        \begin{equation}
            p' = 2 \left[\frac{|\mathrm{Tr} \hat{\Pi}_s|^2}{D_{\textrm{tb}}^2 - 1} - \frac{\mathrm{Tr} (\hat{\Pi}_s^{\dagger} \hat{\sigma}_{\textrm{t}}^x \hat{\Pi}_s \hat{\sigma}_{\textrm{t}}^x)}{D_{\textrm{tb}} (D_{\textrm{tb}}^2 - 1)}\right] = \frac{D_{\textrm{tb}}^2/2 - r_x^2}{D_{\textrm{tb}}^2 - 1}
        \end{equation}

    \section{Noisy Iterative Recovery Schemes} \label{app: error}

        \subsection{Noisy Iterative Scrambling Recovery Protocol} 

        In the noisy scrambling protocol, the fixed point of the recursive equation will be affected by the error occurring during the scrambling procedure.
        The noisy iterative scheme is shown in Fig.~\ref{fig: iterate_error}, where the error in the $n$-th step of scrambling is described as depolarizing error with depolarizing rate $(1-x^{(n)})$
        \begin{equation}
            \mathcal{E}^{(n)}(\rho_{\textrm{tb}})=x^{(n)}\rho_{\textrm{tb}}+(1-x^{(n)})\hat{\mathbb{I}}_{\textrm{tb}}/D_{\textrm{tb}}.
        \end{equation}
        The noises $\mathcal{E}^{(n)}$ in the scrambling should be distinguished from the perturbation $\Lambda^{(n)}$ in $n$-step.
        The perturbations $\Lambda^{(n)}$ act non-trivially only on the target system, which is defined by the original perturbation $\Lambda^{(0)} = \Lambda$, while the noises $\mathcal{E}^{(n)}$ act on the whole system where scrambling occurring.
        
        The total noisy perturbation $\tilde{\Lambda}^{(n)}$ in $n$-th step is the composition of the twirling perturbation $\Lambda_{\mathrm{twirl}}^{(n-1)}$ in $(n-1)$-th step and the noise $\mathcal{E}^{(n-1)}$, which reads as
        \begin{align} \label{eq: perturbation_noise}
            \tilde{\Lambda}^{(n)}(\rho_{\textrm{tb}}) = & \ \Lambda_{\mathrm{twirl}}^{(n-1)} \circ \mathcal{E}^{(n-1)} (\rho_{\textrm{tb}}) \nonumber \\
            = & \ x^{(n-1)} \Lambda_{\mathrm{twirl}}^{(n-1)}(\rho_{\textrm{tb}}) + (1-x^{(n-1)})  \hat{\mathbb{I}}_{\textrm{tb}}/D_{\textrm{tb}} \\
            = & \ x^{(n-1)} \Lambda_{\mathrm{twirl}}^{(n-1)}(\rho_{\textrm{tb}})  + \frac{1-x^{(n-1)}}{D_{\textrm{tb}}^2} \sum_{i=0}^{D_{\textrm{tb}}^2-1} \hat{U}_{i} \rho_{\textrm{tb}} \hat{U}_{i}^{\dagger}, \nonumber
        \end{align}
        where $\hat{U}_i$ are orthogonal unitary basis satisfying $\hat{U}_{0} = \hat{I}$, and $\mathrm{Tr}(\hat{U}_{i}^{\dagger} \hat{U}_{j}) = D_{\mathrm{tb}} \delta_{ij}$.
        Here, the twirling perturbation $\Lambda_{\mathrm{twirl}}^{(k)}$ of $k$-th step is 
        \begin{equation}
            \Lambda_{\mathrm{twirl}}^{(k)}(\cdot) = p^{(k)}\hat{U}_{0} \cdot \hat{U}_{0}^{\dagger} + \frac{1- p^{(k)}}{D_\textrm{t}^2} \sum_{i=0}^{D_{\textrm{t}}^2-1} \hat{U}_{i} \cdot \hat{U}_{i}^{\dagger},
        \end{equation}
        where the $D_{\textrm{t}}^2$ number $\hat{U}_{i}$ are act trivially on the bath of the $(k+1)$-th step of scrambling. 
        Obviously, the perturbation is unital $\Lambda_{\mathrm{twirl}}^{(k)}(\hat{\mathbb{I}}_{\textrm{tb}}) = \hat{\mathbb{I}}_{\textrm{tb}}$, which gives the second term in the second line of Eq.~(\ref{eq: perturbation_noise}).

        With Eq.~(\ref{eq: recovery}), only terms involving $\hat{U}_{0}$ has contribution in recovery rate, by the properties of orthogonal unitary basis.
        The recovery rate of the $n$-step twirling perturbation channel $\Lambda_{\mathrm{twirl}}^{(n)}$ after $n$-th step of scrambling is
        \begin{equation}
            p^{(n)} = (1 - s) x^{(n-1)} p^{(n-1)} + s x^{(n-1)},
        \end{equation}
        where $s = {(D_{\textrm{tb}}^2/D_{\textrm{t}}^2 - 1)}/{(D_{\textrm{tb}}^2 - 1)}$.
        For convenience, let the error channel be the same $\mathcal{E}^{(n)} = \mathcal{E}$ with $x^{(n)}=x$ and from the iterated equation of $p^{(n)}$. 
        We have that
        \begin{equation}
            p^{(n)} = \frac{s x}{1 - (1 - s)x} - \left[(1 - s)x\right]^{n-1} \left(\frac{s x}{1 - (1 - s)x} - \tilde{p}\right),
        \end{equation}
        where $\tilde{p}$ is the recovery rate of the noisy perturbation $\tilde{\Lambda}^{(1)} = \Lambda \circ\mathcal{E}^{(0)}$.
        The recursive formula is still a linear equation, and the factor $(1 - s)x<1$ is less than one, which determines the fixed point as
        \begin{equation}
            p^{(n)} \rightarrow \frac{s x}{1 - (1 - s)x}.
        \end{equation}
        
        When the dimension of bath goes to infinity $D_{\textrm{b}} \rightarrow \infty$, $s \rightarrow {1}/{D_{\textrm{t}}^2}$, we obtain that
        \begin{equation} \label{eq: saturate}
            p^{(n)} \rightarrow \frac{x}{D_{\textrm{t}}^2 + (D_{\textrm{t}}^2-1) x} \approx 1 - D_{\textrm{t}}^2 (1-x) + O[(1-x)^2].
        \end{equation}
        Thus, $p^{(n)}$ scales as $\sim {x}/{D_{\textrm{t}}^2}$, when the error rate $1-x$ is independent of $D_{\textrm{t}}$, which is not better than the original protocol.
        If the error rate scales as $1 - x = \alpha D_{\textrm{t}}^{-2}$, the recovery rate approach to a constant $p^{(\infty)} \rightarrow {1}/{(1 + \alpha)}<1$ as $D_{\textrm{t}} \rightarrow \infty$.
        Since the noisy recursive formula is still linear, the fixed point may move, but not vanish under the influence of noise.
        This demonstrates that the iterative scrambling scheme is robust with the noise in scrambling.

        \begin{figure}[t]
            \centering
            \includegraphics[width=10cm]{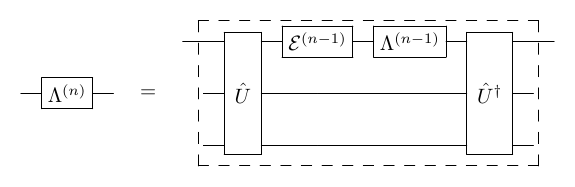}
            \caption{Quantum circuit for the noisy iterative twirling channel.}
            \label{fig: iterate_error}
        \end{figure}

        \subsection{Noisy Iterative ICO-scrambling Scheme} 

        We mainly focus on the noise on the control qubit, because the noise in the scrambling have no substantial difference with the noisy iterative scrambling scheme.
        The depolarizing error on the control qubit with an error rate $(1 - x)$ transforms the output state as
        \begin{equation}
            \rho^{\epsilon(n)}_{\textrm{ctb}} = x \rho^{(n)}_{\textrm{ctb}} + (1- x) {\hat{\mathbb{I}}_{\textrm{ctb}}}/({D_{\textrm{ctb}}}) ,
        \end{equation}
        where $\rho^{(n)}_{\textrm{ctb}}$ is
        shown in Eq.~(\ref{eq: ico_iterative}) and (\ref{eq: ico_coherent_iterative}).
        By performing the post-selection, the recursive equation becomes
        \begin{equation}
            p_{\textrm{eICO}}^{(n)} = \frac{x \left[2 p_{\textrm{eICO}}^{(n-1)} + \frac{D_{\textrm{tb}}^2 - 2}{D_{\textrm{tb}}^2(D_{\textrm{tb}}^2 - 1)}(1 - p_{\textrm{eICO}}^{(n-1)})\right]}{1 + x \left[p_{\textrm{eICO}}^{(n-1)} + \frac{2}{D_{\textrm{tb}}^2}(1 - p_{\textrm{eICO}}^{(n-1)})\right]}.
        \end{equation}
        It is complex to solve this equation, so we only consider the case $D_{\textrm{tb}}=2$, and the equation is simplified as
        \begin{equation}
            p_{\textrm{eICO}}^{(n)} = \frac{4 x (2 p_{\textrm{eICO}}^{(n-1)} + 1)}{3 \left[2 + x (1 + p_{\textrm{eICO}}^{(n-1)})\right]},
        \end{equation}
        with a fixed point
        \begin{equation}
            p_{\textrm{eICO}}^{(\infty)} = \frac{5}{6} - \frac{1}{x} \pm \frac{\sqrt{73 x^2 - 60 x + 36}}{6 x} < x,
        \end{equation}
        when $x \in [0, 1]$.
        Thus, the recovery rate scheme is limited by the noises on the control qubits.
        Moreover, since the fixed point equation derived from the iterative equation is of second order, it is possible that there is no real solution.
        For instance, when $73 x^2 - 60 x + 36 \leq 0$ for the case $D_{\textrm{tb}}=2$.
        This illustrates that there is a threshold for recovery by ICO-scrambling in the noise on the control qubits.

    \section{Quantum simulation of Haar randomness with Unitary $t$-design} \label{app: design}

        Unitary $t$-design of Haar measure on a unitary group $U$ is a finite subset $S$, satisfying that
        \begin{equation}
            \frac{1}{|S|} \sum_{S} f(S,S^{\dagger}) = \int \!\! d\hat{U}\; f(\hat{U}, \hat{U}^{\dagger}),
        \end{equation}
        where $f(\hat{U}, \hat{U}^{\dagger})$ is a homogeneous polynomial with $t$ degree of elements of $\hat{U}$ and also polynomial of elements of $\hat{U}^{\dagger}$.

        Using the unitary design with proper $t$, we can mimic the randomness easily.
        In the case of information scrambling, the random Haar measure is integrated on the so-called twirling channel
        \begin{equation}
        \Lambda_t = \int \!\!d\mathcal{U}\; f(\mathcal{U}, \Lambda) = \int \!\! d\mathcal{U}\; \mathcal{U} \circ \Lambda \circ \mathcal{U}^{\dagger},
        \end{equation}
        whose integrand is a polynomial with $2$-degree of $\hat{U}$ and also $\hat{U}^{\dagger}$.
        So if we average on the unitary $2$-design can mimic the Haar measure theoretically.

        \begin{figure}[ht]
            \centering
            \includegraphics[width=0.48\textwidth]{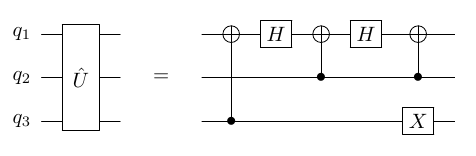}
            \caption{Quantum circuit for the approximate scrambling evolution.}
            \label{fig: evolution}
        \end{figure}

        The method to mimic the Haar randomness used in the original protocol \cite{PhysRevLett.125.040605} is the 3-qubit unitary operator
        \begin{equation}
                \hat{U} = \hat{\mathbb{I}}_{1} \otimes \ket{01} \bra{00} + \hat{\sigma}_1^x \otimes \ket{00} \bra{01}  - \imath \hat{\sigma}_1^y \otimes \ket{11} \bra{10} - \hat{\sigma}_1^z \otimes \ket{10} \bra{11},
        \end{equation}
        whose circuit is shown in Fig. \ref{fig: evolution}.
        Here $q_1$ is the working qubit, $q_2,q_3$ are the bath with initial state $\ket{+}$.
        With this circuit, we can calculate that for any channel $\Lambda$, the corresponding channel twirled by $U$ can send any input of $q_1$ with $\rho$ into the state with reduced density matrix on $q_1$ as
        \begin{equation}
            \rho_{\mathrm{twirl}} = \Lambda_{\mathrm{twirl}}(\rho)
            = \sum_i \sum_k\hat{\sigma}_i \hat{M}_k \hat{\sigma}_i \rho\hat{\sigma}_i \hat{M}_k^{\dagger} \hat{\sigma}_i ,
        \end{equation}
        where $\hat{\sigma}_i \in P = \{\hat{\mathbb{I}}_1, \hat{\sigma}_1^x, \hat{\sigma}_1^y, \hat{\sigma}_1^z\}$ is elements of Pauli group, and $\Lambda(\rho) = \sum_k \hat{M}_k \rho \hat{M}_k^{\dagger}$.
        For Pauli group is only the unitary $1$-design, this can not mimic the twirling channel only, which has integrated with $2$ degree polynomial.
        Thus, it also needs one more average of perturbation $\Lambda$ on Pauli group, which prevent us from manipulating the direction of perturbation.
        Moreover, the 3-qubit unitary is hard to realize when we also want to manipulate its evolution.

        Theoretically, this unitary operator is just a realization of average on unitary $1$-design, we can actually average it "by hand" like the average of the direction of perturbation on Pauli group.
        With this idea, we come back to the average on unitary $t$-design, at less $t = 2$, which have a famous one, the Clifford group (table \ref{tab: Clifford}), and we arrive the circuit (Fig. \ref{fig: design}) in mean text.

        \begin{table}
            \centering
            \begin{tabular}{ccl} \hline\hline
                Axis & Angle & Realization \\ \hline
                    &   &   I \\
                $(1,0,0)$ & $\pi$ & X \\
                $(0,1,0)$ & $\pi$ & Y \\
                $(0,0,1)$ & $\pi$ & Z \\ \hline
                $(1,0,0)$ & $\pi/2$ & X/2 \\
                $(-1,0,0)$ & $\pi/2$ & -X/2 \\
                $(0,1,0)$ & $\pi/2$ & Y/2 \\
                $(0,-1,0)$ & $\pi/2$ & -Y/2 \\
                $(0,0,1)$ & $\pi/2$ & -X/2 Y/2 X/2 \\
                $(0,0,-1)$ & $\pi/2$ & -X/2 -Y/2 X/2 \\ \hline
                $(1,0,1)$ & $\pi$ & X -Y/2 \\
                $(-1,0,1)$ & $\pi$ & X Y/2 \\
                $(0,1,1)$ & $\pi$ & Y X/2 \\
                $(0,-1,1)$ & $\pi$ & Y -X/2 \\
                $(1,1,0)$ & $\pi$ & X/2 Y/2 X/2 \\
                $(1,-1,0)$ & $\pi$ & -X/2 Y/2 -X/2 \\ \hline
                $(1,1,1)$ & $2\pi/3$ & Y/2 X/2 \\
                $(-1,1,1)$ & $2\pi/3$ & Y/2 -X/2 \\
                $(1,-1,1)$ & $2\pi/3$ & -Y/2 X/2 \\
                $(-1,-1,1)$ & $2\pi/3$ & -Y/2 -X/2 \\
                $(1,1,1)$ & $-2\pi/3$ & -X/2 -Y/2 \\
                $(-1,1,1)$ & $-2\pi/3$ & X/2 -Y/2 \\
                $(1,-1,1)$ & $-2\pi/3$ & -X/2 Y/2 \\
                $(-1,-1,1)$ & $-2\pi/3$ & -X/2 -Y/2 \\ \hline\hline
            \end{tabular}
            \caption{Twenty-four  elements of the Clifford group.}
            \label{tab: Clifford}
        \end{table}

    \section{Details of Simulation} \label{app: simulation}

        The experiments are demonstrated on the ScQ-P18 backend of the Quafu platform.
        The layout of the ScQ-P18 device is shown in Fig.~\ref{fig: chip_info}.
        The qubits $q_0,q_1,q_2$ of the circuits in Fig. \ref{subfig: design_ico}, \ref{subfig: design_direction} are mapped to the physical qubits $q_2,q_3,q_4$ of ScQ-P20.
        The qubits $q_0,q_1,q_2,q_3$ of the circuits in Fig. \ref{subfig: design_direction_ico} are mapped to the physical qubits $q_2,q_3,q_4,q_5$ of ScQ-P18.
        The fidelities of CZ gate between them are $0.9874, 0.9934, 0.9821$.
        More information about the four used qubits can be found in Table~\ref{tab: paras}.
        The data and the software of the simulations and experiments are available in the Ref.~\cite{dataset}.

        \begin{figure}[t]
            \centering
            \includegraphics[width=0.6\textwidth]{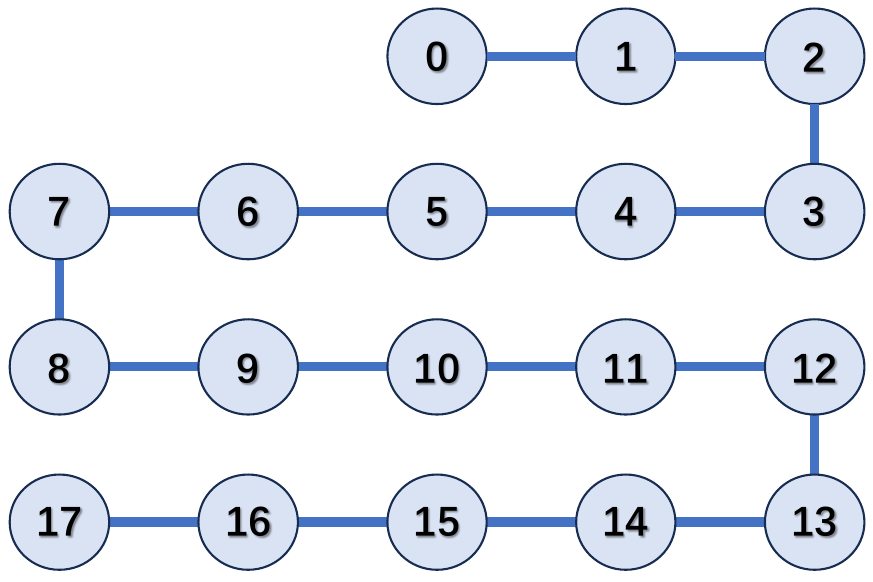}
            \caption{Diagram of ScQ-P18 processor of the on-cloud \emph{Quafu} quantum computers.}
            \label{fig: chip_info}
        \end{figure}

        \begin{table}[b]
            \begin{ruledtabular}
            \centering
            \begin{tabular}{ccccc}
                Qubit index & 2 & 3 & 4 & 5 \\[.06cm] \hline
                    \\[-.4cm]
                    Qubit frequency, $f_{10}$ ($\mathrm{GHz}$) & 4.566 & 4.977 & 4.500 & 4.943 \\[.06cm]
                    Readout frequency, $f_{r}$ ($\mathrm{GHz}$) & 6.737 & 6.713 & 6.693 & 6.675 \\[.06cm]
                    Anharmonicity, $\eta$ ($\mathrm{MHz}$) & -202.6 & -197.9 & -204.8 & -193.7 \\[.06cm]
                    Relaxation time, $T_{1}$ ($\mu\mathrm{s}$) & 41.4 & 40.1 & 29.0 & 36.9 \\[.06cm]
                    Coherence time, $T_{2}$ ($\mu\mathrm{s}$) & 4.83 & 2.30 & 5.69 & 2.05 \\[.06cm] 
                    Single-qubit gate fidelity, $F_{sq}$ & 0.9978 & 0.9936 & 0.9907 & 0.9936 \\[.06cm]
                    Readout fidelity, $F_0$ & 0.9498 & 0.9591 & 0.9405 & 0.9579 \\ [.06cm]
                    Readout fidelity, $F_1$ & 0.8583 & 0.8549 & 0.8781 & 0.8750 \\ [.06cm]
                \end{tabular}
            \end{ruledtabular}
            \caption{The parameters of the four used qubits.} \label{tab: paras}
        \end{table}

        \subsection{The Theoretical Calculation}
            \subsubsection{ICO scheme} \label{subapp: ico}

            When $D_{\textrm{tb}} = D_{\textrm{t} }= 2$, we cannot use the asymptotic result of output in equation (\ref{eq: ico})
            For this case, the non-diagonal state is
            \begin{align}
                    \tilde{\rho}_{\textrm{t}} = & \ \left(p + \frac{1}{D_{\textrm{t}}^2 -1}\right) \rho_{\textrm{t}} + \frac{1}{D_{\textrm{t}}^2 - 1} \sum_k \hat{M}_k^{\dagger} \rho_{\textrm{t}} \hat{M}_k - \frac{1}{D_{\textrm{t}}(D_{\textrm{t}}^2 - 1)} \sum_k [\mathrm{Tr}(\hat{M}_k) \rho_{\textrm{t}} \hat{M}_k^{\dagger} + \mathrm{Tr}(\hat{M}_k^{\dagger}) \hat{M}_k \rho_{\textrm{t}}] \nonumber\\
                    = & \ \frac{1}{3} \rho_{\textrm{t}} + \frac{1}{3} \sum_k \hat{M}_k^{\dagger} \rho_{\textrm{t}} \hat{M}_k
                    = \frac{1}{2} \rho_{\textrm{t}} + \frac{1}{6} \hat{\sigma}_{\textrm{t}}^{\bm{r}} \rho_{\textrm{t}} \hat{\sigma}_{\textrm{t}}^{\bm{r}},
            \end{align}
            and $\langle \hat{\sigma}_{\textrm{c}}^x \rangle = \mathrm{Tr} \tilde{\rho}_{\textrm{t}} = \frac{2}{3}$.
            Thus, we get the output state recovering information as
            \begin{equation}
                \rho_{\textrm{t}}^+ = \frac{1}{2} \rho_{\textrm{t}} + \frac{1}{10} \hat{\sigma}_{\textrm{t}}^{\bm{r}} \rho_{\textrm{t}} \hat{\sigma}_{\textrm{t}}^{\bm{r}} + \frac{2}{5} \frac{\hat{\mathbb{I}}}{D_{\textrm{t}}},
            \end{equation}
            and the fidelity $F = {(7 + r_z^2)}/{10}$, where $r_z = \cos \theta$ is the component of vector ${\bm{r}}$ in $z$ direction.

            \subsubsection{MDD scheme} \label{subapp: direction}

            In this situation, the assumption the dimension $D_{\textrm{tb}} \rightarrow \infty$ goes to infinity fails when we use the Clifford group to mimic the Haar measure of a single qubit.
            Without the asymptotic assumption, the result of output state is
            \begin{equation}
                \rho_{\textrm{atb}} = \frac{1}{2}\hat{\mathbb{I}}_{\textrm{a}} \otimes [p \rho_{\textrm{tb}} + (1 - p) \hat{\mathbb{I}}/D_{\textrm{tb}}]
                + \frac{1}{2} \hat{\sigma}_{\textrm{a}}^x \otimes [p' \rho_{\textrm{tb}} + (r_x^2 - p') \hat{\mathbb{I}}/D_{\textrm{tb}}],
            \end{equation}
            where $p = {1}/{3}, p' = {(2 - r_x^2)}/{3}$ by $D_{\textrm{tb}} = D_{\textrm{t}} = 2$.
            For we only concern about the fidelity of state with $\ket{+}$ in control qubit $q_{\textrm{a}}$, the corresponding state is
            \begin{equation}
                \rho_{\textrm{t}}^{+} = \frac{3 - r_x^2}{3(1 + r_x^2)} \rho_{\textrm{t}} + \frac{2 r_x^2}{3(1 + r_x^2)} {\hat{\mathbb{I}}_{\textrm{t}}},
            \end{equation}
            and the fidelity with initial state $\ket{0}$ is $F =  {2}/{3(1 + r_x^2)} + {1}/{3}$.

            \subsubsection{Combination of ICO and MDD schemes}

            For the diagonal part of $q_{\textrm{a}}$, the states of $q_{\textrm{c}},q_{\textrm{t}}$ is same as the case only with the ICO scheme, thus $\langle \hat{\sigma}_{\textrm{c}}^x \rangle = \mathrm{Tr} \tilde{\rho}_{\textrm{t}} = {2}/{3}$, and the state of $q_{\textrm{t}}$ after post-selection on $q_{\textrm{c}}$ without normalization is
            \begin{equation}
                \rho_{\textrm{t}1}^{+} = \frac{5}{12} \rho_{\textrm{t}} +  \frac{1}{12} \hat{\sigma}_{\textrm{t}}^{\bm{r}} \rho_{\textrm{t}} \hat{\sigma}_{\textrm{t}}^{\bm{r}} + \frac{\hat{\mathbb{I}}_{\textrm{t}}}{6}.
            \end{equation}
            For the non-diagonal part of $q_{\textrm{a}}$, the reduced state of $q_{\textrm{c}},q_{\textrm{t}}$ is
            \begin{align}
                \rho_{\textrm{ct}2} = & \frac{1}{2}\hat{\mathbb{I}}_{\textrm{c}} \otimes [p' \rho_{\textrm{t}} + (r_x^2 - p') \hat{\mathbb{I}}/d] \nonumber\\
                & + \frac{1}{2} \hat{\sigma}^x \otimes [(p + {1}/{6}) \rho_{\textrm{t}} +  \hat{\sigma}_{\textrm{t}}^{\bm{r}} \rho \hat{\sigma}_{\textrm{t}}^{{\bm{r}}'}/6],
            \end{align}
            thus $\langle \hat{\sigma}_{\textrm{a}}^x \rangle = \mathrm{Tr} (\rho_{\textrm{ct}2}) = r_x^2$, and the state of $q_{\textrm{t}}$ without normalization is
            \begin{equation}
                \rho_{\textrm{t}2}^+ = \left(\frac{7}{12} - \frac{r_x^2}{6}\right) \rho_{\textrm{t}} + \frac{1}{12} \hat{\sigma}_{\textrm{t}}^{\bm{r}} \rho_{\textrm{t}} \hat{\sigma}_{\textrm{t}}^{{\bm{r}}'} + \frac{2 r_x^2 - 1}{6} {\hat{\mathbb{I}}_{\textrm{t}}}.
            \end{equation}
            Here we select ${\bm{r}} = (r_x, 0, r_z)$ in circuit, thus ${\bm{r}}' = (r_x, 0, - r_z)$.
            Then we get the normalized state of interest
            \begin{equation}
                \rho_{\textrm{t}}^{++} = \frac{1}{3 + 2 r_x^2} \left[\left(3 - \frac{r_x^2}{2}\right) \rho_{\textrm{t}} + \frac{\hat{\sigma}_{\textrm{t}}^{\bm{r}} \rho_{\textrm{t}} \hat{\sigma}^{{\bm{r}}'} + \hat{\sigma}_{\textrm{t}}^{\bm{r}} \rho_{\textrm{t}} \hat{\sigma}_{\textrm{t}}^{\bm{r}}}{4} +  r_x^2 {\hat{\mathbb{I}}_{\textrm{t}}}\right]
            \end{equation}
            with successful probability $p_{++} = ({2 r_x^2 + 3})/{6}$, and the fidelity $F ={9}/({8 r_x^2 + 12}) + {1}/{4}$.

        \subsection{Simulation of schemes}

        The ICO-scrambling scheme, MDD scheme and their combination are simulated by the unitary 2-design.
        We use the $24$ single-qubit Clifford gates to simulate the scrambling in Haar measure.
        The perturbation is selected as the projective measurement (non-selective).
        The circuits are show in Fig.~\ref{subfig: design_ico}, \ref{subfig: design_direction}, and \ref{subfig: design_direction_ico}.

        Several quantities are of our interest.
        The expectation $\langle \hat{\sigma}_{\textrm{a}}^x \rangle, \langle \hat{\sigma}_{\textrm{c}}^x \rangle$ record the information of the damage.
        The fidelity of target system $F = \mathrm{Tr}(\rho_{\textrm{t}0} \rho_{\textrm{t}})$ evaluates performance of the recovery.
        The successful probability $p_{++} = \mathrm{Tr}(\rho_{\textrm{a}+}\otimes \rho_{\textrm{c}+} \rho_{\textrm{ac}})$ evaluates the overhead of schemes. 
        
        For each of the $24$ Clifford gates, we measure the quantities of interest with $1,000$ shots. 
        The results shown in the main text are uniformly averaged over the $24$ Clifford gate.
        Moreover, the projective measurement directions are under consideration.
        The $50$ directions are uniformly sampled in $x$-$z$ plane with equal spacing.

        \subsection{Error processing} \label{app: fixing}

        To process the error in experiment, we treat the twirling channel as a whole.
        Notice that when perturbation channel $\Lambda = \Lambda^0 \equiv \mathcal{I}$, theoretically the twirling channel of it is also $\Lambda_{\mathrm{twirl}} = \Lambda_{\mathrm{twirl}}^0 \equiv \mathcal{I}$, thus in practical the twirling channel with error is just the error channel $\Lambda_{\mathrm{twirl}}^0 = \mathcal{E}$.
        For general perturbation channel $\Lambda$, the twirling channel with error is $\Lambda_{\mathrm{twirl}}^{e} = \mathcal{E} \circ \Lambda_{\mathrm{twirl}}$.
        So we have to gain adequate knowledge about the error channel $\mathcal{E}$ to deal with the error.
        However, to perform QPT on an operation of $d$ dimension space, there are $d^4 - d^2$ parameters needed to be determined.
        Thus, we want to reduce the parameters needed to process the error.

        In our circuit, the output state of auxiliary or control qubit $q_{\textrm{c}}$ and the target qubit $q_{\textrm{t}}$ after twirling channel $\Lambda_{\mathrm{twirl}}$ can be written as
        \begin{align}
            \rho_{\textrm{ct}} = \Lambda_{\mathrm{twirl}}(\rho_{\textrm{ct}}) =(\hat{\mathbb{I}}_{\textrm{c}} \otimes \rho_{\textrm{t}0} + \hat{\sigma}_{\textrm{c}}^x \otimes \rho_{\textrm{t}1})/2 =  [\ket{+}_{\textrm{c}} \bra{+} \otimes (\rho_{\textrm{t}0} + \rho_{\textrm{t}1}) + \ket{-}_{\textrm{c}} \bra{-} \otimes (\rho_{\textrm{t}0} - \rho_{\textrm{t}1})]/2,
        \end{align}
        where $\rho_{\textrm{t}0},\rho_{\textrm{t}1}$ is normalized state of $q_{\textrm{t}}$.
        Thus, what we need to know is the action of error channel $\mathcal{E}$ on state $\ket{\pm}_{\textrm{c}} \bra{\pm} \otimes \rho_{\textrm{t}}$.
        The joint state space of two qubits $q_{\textrm{c}}$, $q_{\textrm{t}}$ is the tensor product space of state space of $q_{\textrm{c}}$ and $q_{\textrm{t}}$, and the bases of tensor space are the tensor products of the bases of $q_{\textrm{c}}$ and $q_{\textrm{t}}$.
        Selecting the Pauli matrix $\sigma^{\mu}, (\mu = 0, 1, 2, 3)$, as the bases of state space of qubits $q_{\textrm{c}}$ and $q_{\textrm{t}}$, the bases of the tensor space is $\hat{\sigma}_{\textrm{c}}^{\mu} \otimes \hat{\sigma}_{\textrm{t}}^{\nu}$, and the joint state of $q_{\textrm{c}},q_{\textrm{t}}$ can be decomposed as
        \begin{equation}
            \mathcal{E}(\ket{\pm}_{\textrm{c}} \bra{\pm} \otimes \rho_{\textrm{t}}) = \frac{1}{2} \sum_{\mu,\nu} \chi_{\mu \nu}^{\pm} \hat{\sigma}_{\textrm{c}}^{\mu} \otimes \hat{\sigma}_{\textrm{t}}^{\nu} = \frac{1}{2} \sum_{\mu} p_{\mu}^{\pm} \hat{\sigma}_{\textrm{c}}^{\mu} \otimes \rho_{\textrm{t}}^{\mu \pm},
        \end{equation}
        where $p_{\mu}^{\pm} \rho_{\mu}^{\pm} = \sum_{\nu} \chi_{\mu \nu}^{\pm} \hat{\sigma}_{\nu}$, $\rho_{\mu}$ is normalized, and $p_0^{\pm} = 1$ to make sure $\mathcal{E}$ preserving the trace of state.
        Then we denote the completely positive trace preserving (CPTP) operator mapping $\rho$ to $\rho_{\mu}^{\pm}$ as $\mathcal{E}_{\mu}^{\pm}$, we have
        \begin{equation}
            \mathcal{E}(\ket{\pm}_{\textrm{c}} \bra{\pm} \otimes \rho_{\textrm{t}}) = \frac{1}{2} \sum_{\mu} p_{\mu}^{\pm} \hat{\sigma}_{\textrm{c}}^{\mu} \otimes \mathcal{E}_{\mu}^{\pm}(\rho_{\textrm{t}}).
        \end{equation}
        Applying this result to the twirling channel with error, we have
        \begin{equation}
            \rho^{e}_{\textrm{ct}} = \Lambda_{\mathrm{twirl}}^e(\rho_{\textrm{c}}) = \frac{1}{4} \sum_{\mu} \hat{\sigma}_{\textrm{c}}^{\mu} \otimes \left[p_{\mu}^{+} \mathcal{E}_{\mu}^{+}(\rho_{\textrm{t}}^+) + p_{\mu}^{-} \mathcal{E}_{\mu}^{-}(\rho_{\textrm{t}}^-)\right],
        \end{equation}
        where $\rho_{\textrm{t}}^{\pm} = \rho_{\textrm{t}0} \pm \rho_{\textrm{t}1}$.

        When measure $\hat{\sigma}_{\textrm{c}}^{x}$ on $q_{\textrm{c}}$, we consider the reduced density matrix of $q_{\textrm{c}}$,
        \begin{align}
            \rho_{\textrm{c}} & =  \hat{\mathbb{I}}_{\textrm{c}}/2 + q \hat{\sigma}_{\textrm{c}}^{x}/2 , \\
            \rho_{\textrm{c}}^e & = \hat{\mathbb{I}}_{\textrm{c}}/2 + \sum_{i} \hat{\sigma}_{\textrm{c}}^{i} [p_{i}^{+} (1+q) + p_{i}^{-}(1-q)]/4,
        \end{align}
        where $q = \mathrm{Tr} \rho_{\textrm{t}1}$, and $i = 1, 2, 3$.
        For the twirling channel with or without error, we have
        \begin{equation}
            \braket{\hat{\sigma}_{\textrm{c}}^{x}} = q, \qquad
            \braket{\hat{\sigma}_{\textrm{c}}^{x}}_{e} = \frac{1}{2} [p_{1}^{+} (1+q) + p_{1}^{-}(1-q)],
        \end{equation}
            and for the error channel with initial state $\ket{\pm}_{\textrm{c}} \otimes \rho_{\textrm{t}}$,
        \begin{equation}
            \braket{\hat{\sigma}_{\textrm{c}}^{x}}_{\pm} = p_{1}^{\pm}.
        \end{equation}
        So the expectation $\braket{\hat{\sigma}_{\textrm{c}}^{x}}_{e}$ of twirling channel with error can be evaluated by the expectation without error $\braket{\hat{\sigma}_{\textrm{c}}{x}}$ and of the error channel $\braket{\hat{\sigma}_{\textrm{c}}^{x}}_{\pm}$,
        \begin{equation}
            \braket{\hat{\sigma}_{\textrm{c}}^{x}}_{e} = \frac{\braket{\hat{\sigma}_{\textrm{c}}^{x}}_{+} + \braket{\hat{\sigma}_{\textrm{c}}^{x}}_{-}}{2} + \frac{\braket{\hat{\sigma}_{\textrm{c}}^{x}}_{+} - \braket{\hat{\sigma}_{\textrm{c}}^{x}}_{-}}{2} \braket{\hat{\sigma}_{\textrm{c}}^{x}},
        \end{equation}
        where $\braket{\hat{\sigma}_{\textrm{c}}^{x}}$ is the theoretical value, and the $\braket{\hat{\sigma}_{\textrm{c}}^{x}}_{\pm}$ can be measured from the experiment by removing perturbation and setting corresponding initial state on $q_{\textrm{c}}$.

        When calculation the fidelity of $q_{\textrm{t}}$, we should consider the reduced matrix of $q_{\textrm{t}}$, and particularly, we post select the state of $q_{\textrm{t}}$ with $q_{\textrm{c}}$ in $\ket{+}_{\textrm{c}}$ state.
        The reduced matrix of $q_{\textrm{t}}$
        \begin{align}
            \rho_{\textrm{t}}^{e+} =\ & \frac{[(1 + p_1^{+}) \mathcal{E}^{+}  + (1 + p_1^{-}) \mathcal{E}^{-}] (\rho_{\textrm{t}0})}{2 + (p_1^{+} + p_1^{-}) + q (p_1^{+} - p_1^{-})} + \frac{[(1 + p_1^{+}) \mathcal{E}^{+} - (1 + p_1^{-}) \mathcal{E}^{-}](\rho_{\textrm{t}1})}{2 + (p_1^{+} + p_1^{-}) + q (p_1^{+} - p_1^{-})}
        \end{align}
        where $\mathcal{E}^{\pm} =  \left(\mathcal{E}_{0}^{\pm} + p_{1}^{\pm} \mathcal{E}_{1}^{\pm}\right)/(1 + p_{1}^{\pm})$ is the normalized error channel on $q_{\textrm{t}}$ with initial state $\ket{\pm}_{\textrm{c}}$ and post selected by $\ket{+}_{\textrm{c}}$ on $q_{\textrm{c}}$.
        To calculate the fidelity explicitly, we need to consider the circuit in detail.

            \subsubsection{ICO scheme} \label{app: fixing_ico}

            In this scheme, as we have shown previously,
            \begin{equation}
                \rho_{\textrm{t}0} = \frac{1}{3} \rho_{\textrm{t}} + \frac{1}{3} \hat{\mathbb{I}}_{\textrm{t}}, \qquad
                \rho_{\textrm{t}1} = \frac{1}{2} \rho_{\textrm{t}} + \frac{1}{6} \hat{\sigma}_{\textrm{t}}^{\bm{r}} \rho_{\textrm{t}} \hat{\sigma}_{\textrm{t}}^{\bm{r}},
            \end{equation}
            and $q ={2}/{3}$.
            So the problem reduce to calculate the fidelities between $\mathcal{E}^{\pm}(\rho_{\textrm{t}})$ or $\mathcal{E}^{\pm}(\hat{\sigma}_{\textrm{t}}^{\bm{r}} \rho_{\textrm{t}} \hat{\sigma}_{\textrm{t}}^{\bm{r}})$ and $\rho_{\textrm{t}}$, where $\rho_{\textrm{t}}$ is a pure state.

            For $\mathcal{E}^{\pm}(\rho_{\textrm{t}})$, the fidelity $F^{\pm} = \mathrm{Tr}[\rho_{\textrm{t}} \mathcal{E}^{\pm}(\rho_{\textrm{t}})]$ can be set as the average fidelity of channel $\mathcal{E}^{\pm}$, which can be measured by removing the perturbation in our circuit.
            For $\mathcal{E}^{\pm}(\hat{\sigma}_{\textrm{t}}^{\bm{r}} \rho_{\textrm{t}} \hat{\sigma}_{\textrm{t}}^{\bm{r}})$ with $\hat{\sigma}_{\textrm{t}}^{\bm{r}} \equiv {\bm{r}} \cdot \hat{\sigma}_{\textrm{t}}$ and $\rho_{\textrm{t}} = |0\rangle _{\textrm{t}}\langle0|$, the fidelity
            \begin{align}
                \mathrm{Tr}[\rho_{\textrm{t}} \mathcal{E}^{\pm}(\hat{\sigma}_{\textrm{t}}^{\bm{r}} \rho_{\textrm{t}} \hat{\sigma}_{\textrm{t}}^{\bm{r}})]
                =&  \ \braket{0| \mathcal{E}^{\pm}(\hat{\sigma}_{\textrm{t}}^{\bm{r}} |0\rangle _{\textrm{t}}\langle0| \hat{\sigma}_{\textrm{t}}^{\bm{r}})|0}\nonumber\\
                =& \sum_{i,j = 0,1} \bra{i| \hat{\sigma}_{\textrm{t}}^{\bm{r}} |0} \braket{0| \hat{\sigma}_{\textrm{t}}^{\bm{r}} |j} \braket{0| \mathcal{E}^{\pm}(|i\rangle _{\textrm{t}}\langle j|)|0} \\
                = & \  \left|\braket{0| \hat{\sigma}_{\textrm{t}}^{\bm{r}} |0}\right|^2 \braket{0| \mathcal{E}^{\pm}(|0\rangle _{\textrm{t}}\langle0|)|0} + \left|\braket{1| \hat{\sigma}_{\textrm{t}}^{\bm{r}} |0}\right|^2 \braket{0| \mathcal{E}^{\pm}(|1\rangle _{\textrm{t}}\langle1|)|0} \nonumber\\
                & + \braket{0| \hat{\sigma}_{\textrm{t}}^{\bm{r}} |0} \braket{0| \hat{\sigma}_{\textrm{t}}^{\bm{r}} |1} \braket{0| \mathcal{E}^{\pm}(|0\rangle _{\textrm{t}}\langle1|)|0}+ \braket{1| \hat{\sigma}_{\textrm{t}}^{\bm{r}} |0} \braket{0| \hat{\sigma}_{\textrm{t}}^{\bm{r}} |0} \braket{0| \mathcal{E}^{\pm}(|1\rangle _{\textrm{t}}\langle0|)|0}.
            \end{align}
            With  ${\bm{r}} = (r_x, 0, r_z)$, we have $\braket{0| \hat{\sigma}_{\textrm{t}}^{\bm{r}} |0} = r_z, \braket{0| \hat{\sigma}_{\textrm{t}}^{\bm{r}} |1} = \braket{1| \hat{\sigma}_{\textrm{t}}^{\bm{r}} |0} = r_x$.
            As before, we can set $\braket{0| \mathcal{E}^{\pm}(|0\rangle _{\textrm{t}}\langle0|)|0} = F^{\pm}$, then $\braket{0| \mathcal{E}^{\pm}(|1\rangle _{\textrm{t}}\langle1|)|0} = 1 - F^{\pm}$ can be verified.
            We can also assume that the error channel $\mathcal{E}^{\pm}$ is Hermitian, for the circuit of $\Lambda_{t}^0 = \mathcal{E}^{\pm}$ is invariant under time reversing.
            With above results, we can represent the fidelity as
            \begin{equation}
                \mathrm{Tr}[\rho_{\textrm{t}} \mathcal{E}^{\pm}(\hat{\sigma}_{\textrm{t}}^{\bm{r}} \rho_{\textrm{t}} \hat{\sigma}^{\bm{r}})]
                = r_z^2 F^{\pm} + r_x^2 (1 - F^{\pm}) + r_x r_z \mathrm{Tr}[\hat{\sigma}_{\textrm{t}}^x \mathcal{E}^{\pm}(|0\rangle _{\textrm{t}}\langle0|)],
            \end{equation}
            and the problem is to calculate the polarization of state after error channel.

            In general, the state after error channel is not a pure state and is expressed as
            \begin{equation}
                \mathcal{E}^{\pm}(|0\rangle _{\textrm{t}}\langle0|) = F^{\pm} |0\rangle _{\textrm{t}}\langle0| + (1- F^{\pm}) |0\rangle _{\textrm{t}}\langle0| + r (e^{\imath \theta}|0\rangle _{\textrm{t}}\langle1| + e^{-\imath \theta}|1\rangle _{\textrm{t}}\langle0|).
            \end{equation}
            Then, we have $\mathrm{Tr}[\hat{\sigma}_x \mathcal{E}^{\pm}(|0\rangle _{\textrm{t}}\langle0|)] = 2 r \cos \theta$.
            Because our circuit is average over Clifford group, in this average,
            \begin{equation}
                \langle\cos \theta\rangle = \frac{1}{N} \sum_{\phi_i} \cos (\phi_i + \theta) \approx \int\!\! d\phi\; \cos (\phi + \theta) = 0,
            \end{equation}
            the polarization $\mathrm{Tr}[\hat{\sigma}_x \mathcal{E}^{\pm}(|0\rangle _{\textrm{t}}\langle0|)] = 0$, and
            \begin{equation}
                \mathrm{Tr}[\rho \mathcal{E}^{\pm}(\hat{\sigma}_{\textrm{t}}^{\bm{r}} \rho_{\textrm{t}} \hat{\sigma}_{\textrm{t}}^{\bm{r}})] = r_z^2 F^{\pm} + r_x^2 (1 - F^{\pm}).
            \end{equation}
            Then the fidelities of $\rho_{\textrm{t}0}, \rho_{\textrm{t}1}$ with respect to $\rho_{\textrm{t}}$ are
            \begin{align}
                F_{0}^{\pm} & = \frac{1}{3} F^{\pm} + \frac{1}{3}, \\
                F_{1}^{\pm} & = \frac{2 - r_x^2}{3} F^{\pm} + \frac{1}{6} r_x^2 .
            \end{align}
            The fidelity of twirling channel with error and the expectation $\braket{\hat{\sigma}_{\textrm{c}}^{x}}_e$ are
            \begin{align}
                F^{e} = & \ \frac{(1 + \braket{\hat{\sigma}_{\textrm{c}}^{x}}_{+})(1 + F^{+}) + (1 + \braket{\hat{\sigma}_{\textrm{c}}^{x}}_{-})(1 + F^{-})}{6 + 3(\braket{\hat{\sigma}_{\textrm{c}}^{x}}_{+} + \braket{\hat{\sigma}_{\textrm{c}}^{x}}_{-}) + 2 (\braket{\hat{\sigma}_{\textrm{c}}^{x}}_{+} - \braket{\hat{\sigma}_{\textrm{c}}^{x}}_{-})}  + \frac{(2 - r_x^2) [(1 + \braket{\hat{\sigma}_{\textrm{c}}^{x}}_{+}) F^{+} - (1 + \braket{\hat{\sigma}_{\textrm{c}}^{x}}_{-}) F^{-}]}{6 + 3(\braket{\hat{\sigma}_{\textrm{c}}^{x}}_{+} + \braket{\hat{\sigma}_{\textrm{c}}^{x}}_{-}) + 2 (\braket{\hat{\sigma}_{\textrm{c}}^{x}}_{+} - \braket{\hat{\sigma}_{\textrm{c}}^{x}}_{-})} \nonumber\\
                & \ + \frac{r_x^2}{2} \frac{\braket{\hat{\sigma}_{\textrm{c}}^{x}}_{+} - \braket{\hat{\sigma}_{\textrm{c}}^{x}}_{-}}{6 + 3(\braket{\hat{\sigma}_{\textrm{c}}^{x}}_{+} + \braket{\hat{\sigma}_{\textrm{c}}^{x}}_{-}) + 2 (\braket{\hat{\sigma}_{\textrm{c}}^{x}}_{+} - \braket{\hat{\sigma}_{\textrm{c}}^{x}}_{-})}, \\
                \braket{\hat{\sigma}_{\textrm{c}}^{x}}_e = & \ \frac{\braket{\hat{\sigma}_{\textrm{c}}^{x}}_{+} + \braket{\hat{\sigma}_{\textrm{c}}^{x}}_{-}}{2} + \frac{\braket{\hat{\sigma}_{\textrm{c}}^{x}}_{+} - \braket{\hat{\sigma}_{\textrm{c}}^{x}}_{-}}{3},
            \end{align}
            where $\braket{\hat{\sigma}_{\textrm{c}}^{x}}_{\pm}$ and $F^{\pm}$ are measured from circuit of the twirling channel $\Lambda_{\mathrm{twirl}}^0$ of identical channel $\mathcal{I}$ with corresponding initial states $\ket{\pm}_{\textrm{c}}$ of $q_{\textrm{c}}$ and post-selected on the state $\ket{+}_{\textrm{c}}$.
            The quantities $\braket{\hat{\sigma}_{\textrm{c}}^{x}}_{\pm}$ and $F^{\pm}$ measured from experiments are shown in Fig.~\ref{fig: ico_exp_noise}.

            \begin{figure*}[t]
                \centering
                \subfigure[
                ]{
                    \includegraphics[width=8cm,clip=True]{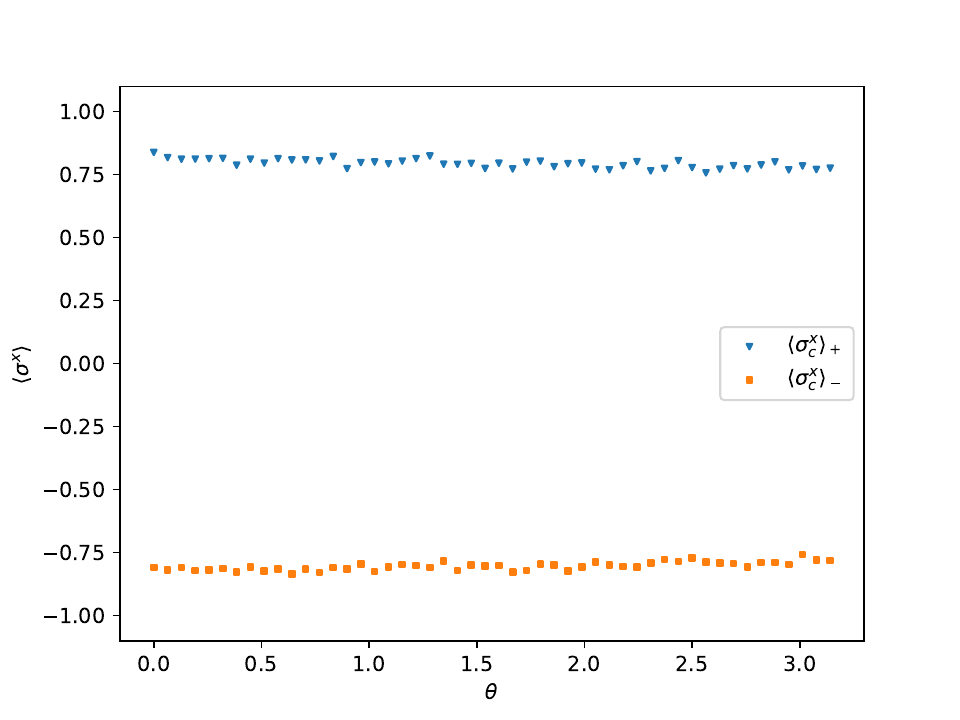}
                    \label{subfig: ico_p_noise}
                }
                \subfigure[
                ]{
                    \includegraphics[width=8cm,clip=True]{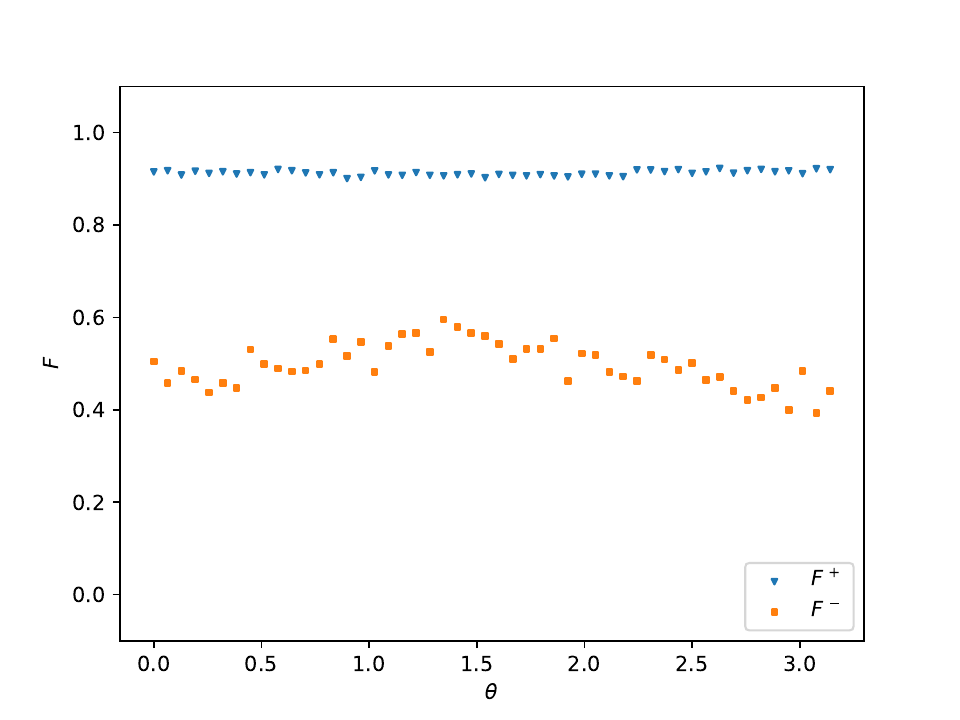}
                    \label{subfig: ico_fidelity_noise}
                }
                \caption{
                    The quantities $\braket{\sigma_c^x}_{\pm}$ and $F^{\pm}$ measured from auxiliary circuits where perturbation absents is show in~\ref{sub@subfig: ico_p_noise} and \ref{sub@subfig: ico_fidelity_noise}, respectively.
                    They are help to predict the theoretical results of measurement outcome in original circuits under the noise of experiments, which have shown in Fig.~\ref{fig: exp_ico}.
                }
                \label{fig: ico_exp_noise}
            \end{figure*}

            \subsubsection{MDD scheme} \label{app: fixing_d}

            In this scheme, as shown in previous,
            \begin{equation}
                \rho_{\textrm{t}0} = \frac{1}{3} \rho_{\textrm{t}} + \frac{2}{6} \hat{\mathbb{I}}_{\textrm{t}}, \qquad
                \rho_{\textrm{t}1} = \frac{2 - r_x^2}{3} \rho_{\textrm{t}} + \frac{2 r_x^2 - 1}{3} \hat{\mathbb{I}}_{\textrm{t}},
            \end{equation}
            and $q = r_x^2$.
            It is also obvious from previous,
            \begin{align}
                F_{0}^{\pm} & = \frac{1}{3} F^{\pm} + \frac{1}{3}, \\
                F_{1}^{\pm} & = \frac{2 - r_x^2}{3} F^{\pm} +\frac{2 r_x^2 - 1}{3},
            \end{align}
            thus the fidelity of twirling channel with error
            \begin{align}
                F^{e} = & \ \frac{1}{3} \frac{(1 + \braket{\hat{\sigma}_{\textrm{a}}^{x}}_{+})(1 + F^{+}) + (1 + \braket{\hat{\sigma}_{\textrm{a}}^{x}}_{-})(1 + F^{-})}{2 + \braket{\hat{\sigma}_{\textrm{a}}^{x}}_{+} + \braket{\hat{\sigma}_{\textrm{a}}^{x}}_{-} + r_x^2 (\braket{\hat{\sigma}_{\textrm{a}}^{x}}_{+} - \braket{\hat{\sigma}_{\textrm{a}}^{x}}_{-})}  + \frac{(2 - r_x^2)}{3} \frac{(1 + \braket{\hat{\sigma}_{\textrm{a}}^{x}}_{+}) F^{+} - (1 + \braket{\hat{\sigma}_{\textrm{a}}^{x}}_{-}) F^{-}}{2 + \braket{\hat{\sigma}_{\textrm{a}}^{x}}_{+} + \braket{\hat{\sigma}_{\textrm{a}}^{x}}_{-} + r_x^2 (\braket{\hat{\sigma}_{\textrm{a}}^{x}}_{+} - \braket{\hat{\sigma}_{\textrm{a}}^{x}}_{-})} \nonumber\\
                & \ + \frac{2 r_x^2 - 1}{3} \frac{\braket{\hat{\sigma}_{\textrm{a}}^{x}}_{+} - \braket{\hat{\sigma}_{\textrm{a}}^{x}}_{-}}{2 + \braket{\hat{\sigma}_{\textrm{a}}^{x}}_{+} + \braket{\hat{\sigma}_{\textrm{a}}^{x}}_{-} + r_x^2 (\braket{\hat{\sigma}_{\textrm{a}}^{x}}_{+} - \braket{\hat{\sigma}_{\textrm{a}}^{x}}_{-})}, \\
                \braket{\hat{\sigma}_{\textrm{a}}^{x}}_e = & \ \frac{\braket{\hat{\sigma}_{\textrm{a}}^{x}}_{+} + \braket{\hat{\sigma}_{\textrm{a}}^{x}}_{-}}{2} + r_x^2 \frac{\braket{\hat{\sigma}_{\textrm{a}}^{x}}_{+} - \braket{\hat{\sigma}_{\textrm{a}}^{x}}_{-}}{2} ,
            \end{align}
            The quantities $\braket{\hat{\sigma}_{\textrm{a}}^{x}}_{\pm}, F^{\pm}$ can be measured similar to ICO scheme, and the results from experiments is shown in Fig.~\ref{fig: d_exp_noise}.

            \begin{figure*}[t]
                \centering
                \subfigure[
                ]{
                    \includegraphics[width=8cm,clip=True]{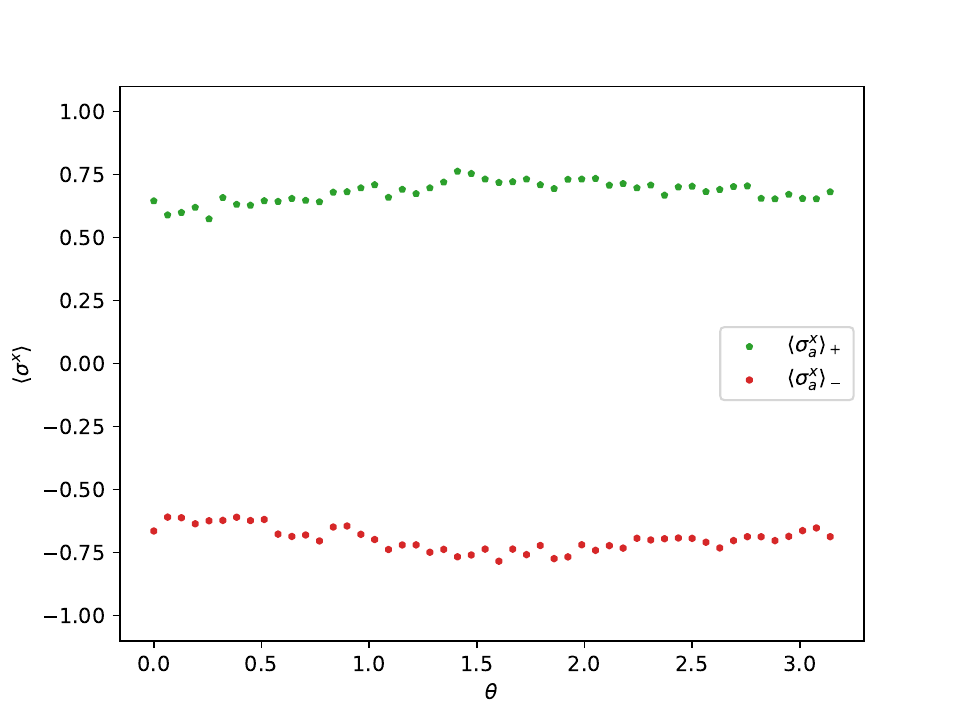}
                    \label{subfig: d_p_noise}
                }
                \subfigure[
                ]{
                    \includegraphics[width=8cm,clip=True]{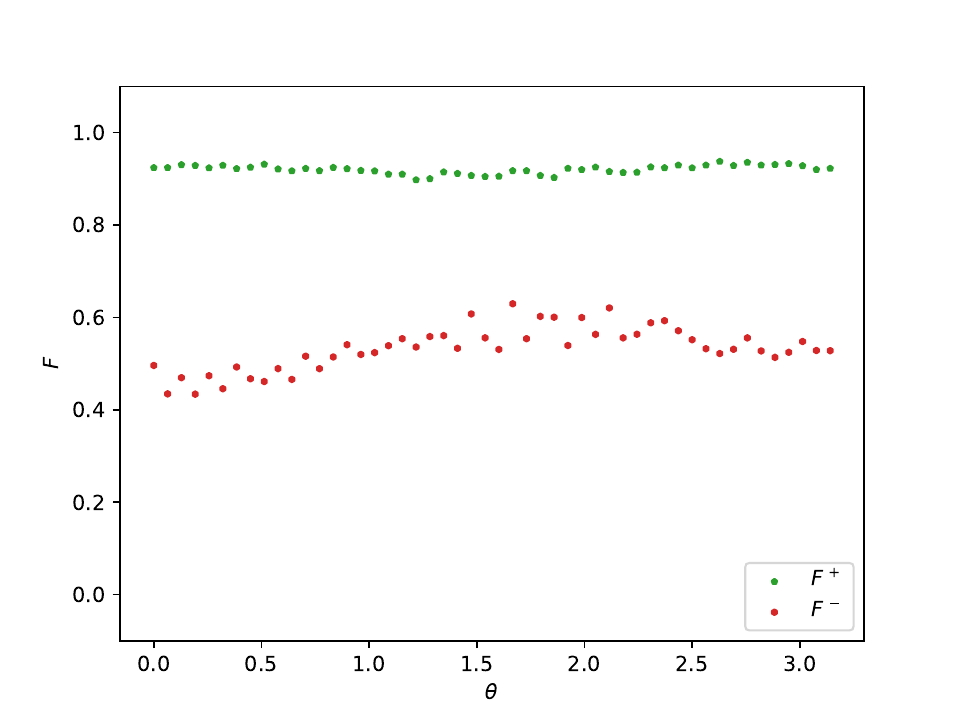}
                    \label{subfig: d_fidelity_noise}
                }
                \caption{
                    The quantities $\braket{\sigma_a^x}_{\pm}$ and $F^{\pm}$ measured from auxiliary circuits where perturbation absents is show in~\ref{sub@subfig: d_p_noise} and \ref{sub@subfig: d_fidelity_noise}, respectively.
                    They are help to predict the theoretical results of measurement outcome in original circuits under the noise of experiments, which have shown in Fig.~\ref{fig: exp_anisotropic}.
                }
                \label{fig: d_exp_noise}
            \end{figure*}

            \subsubsection{Combination of ICO and MDD scheme} \label{app: fixing_ico_d}

            In this scheme, there are two error channel, where $\mathcal{E}_{\textrm{a}}$ is the channel applied on qubits $q_{\textrm{t}}$, $q_{\textrm{a}}$, which same as the one in MDD scheme, and $\mathcal{E}_{\textrm{c}}$, is applied on qubit $q_{\textrm{c}}$, $q_{\textrm{t}}$, which is same as the recovery in ICO scheme.
            Thus, the expectation of $\hat{\sigma}^x$ on control qubit $q_{\textrm{c}}$ and auxiliary qubit $q_{\textrm{a}}$ are same as previous two schemes,
            \begin{align}
                \braket{\hat{\sigma}_{\textrm{c}}^{x}}_e = & \ \frac{\braket{\hat{\sigma}_{\textrm{c}}^{x}}_{+} + \braket{\hat{\sigma}_{\textrm{c}}^{x}}_{-}}{2} + r_x^2 \frac{\braket{\hat{\sigma}_{\textrm{c}}^{x}}_{+} - \braket{\hat{\sigma}_{\textrm{c}}^{x}}_{-}}{2} , \\
                \braket{\hat{\sigma}_{\textrm{a}}^{x}}_e = & \ \frac{\braket{\hat{\sigma}_{\textrm{a}}^{x}}_{+} + \braket{\hat{\sigma}_{\textrm{a}}^{x}}_{-}}{2} + \frac{\braket{\hat{\sigma}_{\textrm{a}}^{x}}_{+} - \braket{\hat{\sigma}_{\textrm{a}}^{x}}_{-}}{3} .
            \end{align}
            The diagonal part of the auxiliary qubit $q_{\textrm{a}}$ is same as the scheme of recovery in ICO, thus the reduced density matrix of $q_{\textrm{t}}$ of $\rho_{\textrm{t}0}$ without normalization is
            \begin{align}
                {\rho}_{\textrm{t}0} = & \ \frac{1}{24} (1 + p_{1\textrm{c}}^{+}) [5 \mathcal{E}_{\textrm{c}}^{+}(\rho_{\textrm{t}}) + \mathcal{E}_{\textrm{c}}^{+}(\hat{\sigma}_{\textrm{t}}^{\bm{r}} \rho_{\textrm{t}} \hat{\sigma}_{\textrm{t}}^{\bm{r}})] - \frac{1}{24} (1 + p_{1\textrm{c}}^{-}) [\mathcal{E}_{\textrm{c}}^{-}(\rho_{\textrm{t}}) + \mathcal{E}_{\textrm{c}}^{-}(\hat{\sigma}_{\textrm{t}}^{\bm{r}} \rho_{\textrm{t}} \hat{\sigma}_{\textrm{t}}^{\bm{r}})] + \frac{1}{12} (2 + p_{1\textrm{c}}^{+} + p_{1\textrm{c}}^{-}) {\hat{\mathbb{I}}_{\textrm{t}}}
            \end{align}
            The coherent part is shown in previous as
            \begin{align}
                \rho_{\textrm{ct}1} = \frac{1}{2} \left(\hat{\mathbb{I}}_{\textrm{c}} \otimes \left[\frac{2 - r_x^2}{3} \rho_{\textrm{t}} + \frac{2(2 r_x^2 - 1)}{6} \hat{\mathbb{I}}_{\textrm{t}}\right] + \hat{\sigma}_{\textrm{c}}^x \otimes \left[\frac{1}{2} \rho_{\textrm{t}} + \frac{1}{6} \hat{\sigma}_{\textrm{t}}^{\bm{r}} \rho_{\textrm{t}} \hat{\sigma}_{\textrm{t}}^{{\bm{r}}'}\right]\right),
            \end{align}
            Then we consider the reduced density matrix of $q_{\textrm{t}}$ after error channel $\mathcal{E}_{c}$, the reduced density matrix of $q_{\textrm{t}}$ without normalization is
            \begin{align}
                {\rho}_{\textrm{t}1} = & \ \frac{1}{24} (1 + p_{1\textrm{c}}^{+}) [(7 - 2 r_x^2) \mathcal{E}_{\textrm{c}}^{+}(\rho_{\textrm{t}}) + \mathcal{E}_{\textrm{c}}^{+}(\hat{\sigma}_{\textrm{t}}^{\bm{r}} \rho_{\textrm{t}} \hat{\sigma}_{\textrm{t}}^{{\bm{r}}'})] + \frac{1}{24} (1 + p_{1\textrm{c}}^{-}) [(1 - 2 r_x^2)\mathcal{E}_{\textrm{c}}^{-}(\rho_{\textrm{t}}) - \mathcal{E}_{\textrm{c}}^{-}(\hat{\sigma}_{\textrm{t}}^{\bm{r}} \rho_{\textrm{t}} \hat{\sigma}_{\textrm{t}}^{{\bm{r}}'})] \nonumber \\
                & + \frac{2 r_x^2 - 1}{12} (2 + p_{1\textrm{c}}^{+} + p_{1\textrm{c}}^{-}) {\hat{\mathbb{I}}_{\textrm{t}}}.
            \end{align}
            Then the reduced state of $q_{\textrm{t}}$, $q_{\textrm{a}}$ by post selection on $q_{\textrm{c}}$ without normalization is
            \begin{equation}
                \rho_{\textrm{ta}} = \frac{1}{2} ({\rho}_{\textrm{t}0} \otimes \hat{\mathbb{I}}_{\textrm{a}} + {\rho}_{\textrm{t}1} \otimes \hat{\sigma}_{\textrm{a}}^x),
            \end{equation}
            and after performing the error channel $\mathcal{E}_{\textrm{a}}$ and post-selecting $q_{\textrm{a}}$, the normalized state can be written as
            \begin{equation}
                \rho_{\textrm{t}}^{++} = \frac{1}{4 p_{++}} \left[\left((1 + p_{1\textrm{a}}^{+}) \mathcal{E}_{\textrm{a}}^{+}  + (1 + p_{1\textrm{a}}^{-}) \mathcal{E}_{\textrm{a}}^{-} \right)({\rho}_{\textrm{t}0}) + \left((1 + p_{1\textrm{a}}^{+}) \mathcal{E}_{\textrm{a}}^{+} - (1 + p_{1\textrm{a}}^{-}) \mathcal{E}_{\textrm{a}}^{-}\right)({\rho}_{\textrm{t}1})\right]
            \end{equation}
            with the probability
            \begin{align}
            p_{++} &=  \ \mathrm{Tr}\rho_{\textrm{t}}^{++}\nonumber\\
            &=  \ \frac{2 + p_{1\textrm{a}}^{+} + p_{1\textrm{a}}^{-}}{16} \left[2 + p_{1\textrm{c}}^{+} + p_{1\textrm{c}}^{-} + \frac{2}{3} (p_{1\textrm{c}}^{+} - p_{1\textrm{c}}^{-})\right]  + \frac{p_{1\textrm{a}}^{+} - p_{1\textrm{a}}^{-}}{16} \left[(2 + p_{1\textrm{c}}^{+} + p_{1\textrm{c}}^{-}) r_x^2 + \frac{1}{3} (r_x^2 + 1) (p_{1\textrm{c}}^{+} - p_{1\textrm{c}}^{-})\right].\nonumber
            \end{align}
            The fidelity of this state with the initial state reduces to the fidelity of $\mathcal{E}_{\textrm{a}}^{\pm} \circ \mathcal{E}_{\textrm{c}}^{\pm}(\rho), \mathcal{E}_{\textrm{a}}^{\pm} \circ \mathcal{E}_{\textrm{c}}^{\pm}(\hat{\sigma}^{\bm{r}} \rho \hat{\sigma}^{\bm{r}})$, and $\mathcal{E}_{\textrm{a}}^{\pm} \circ \mathcal{E}_{\textrm{c}}^{\pm}(\hat{\sigma}^{\bm{r}} \rho \hat{\sigma}^{{\bm{r}}'})$.

            \begin{figure*}[t]
                \centering
                \subfigure[
                ]{
                    \includegraphics[width=8cm,clip=True]{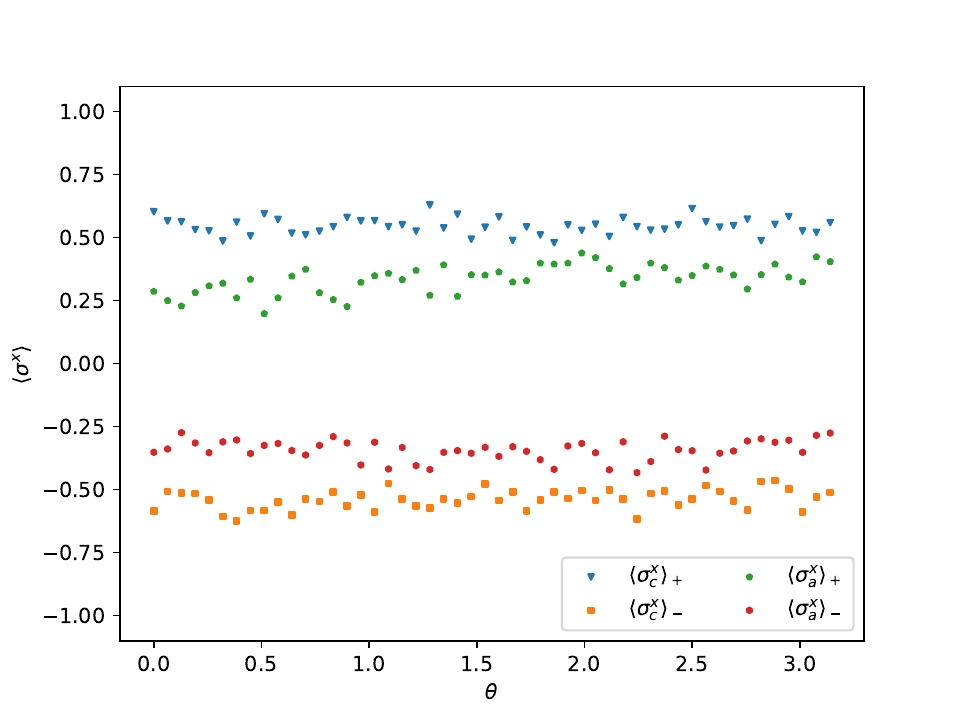}
                    \label{subfig: d_ico_p_noise}
                }
                \subfigure[
                ]{
                    \includegraphics[width=8cm,clip=True]{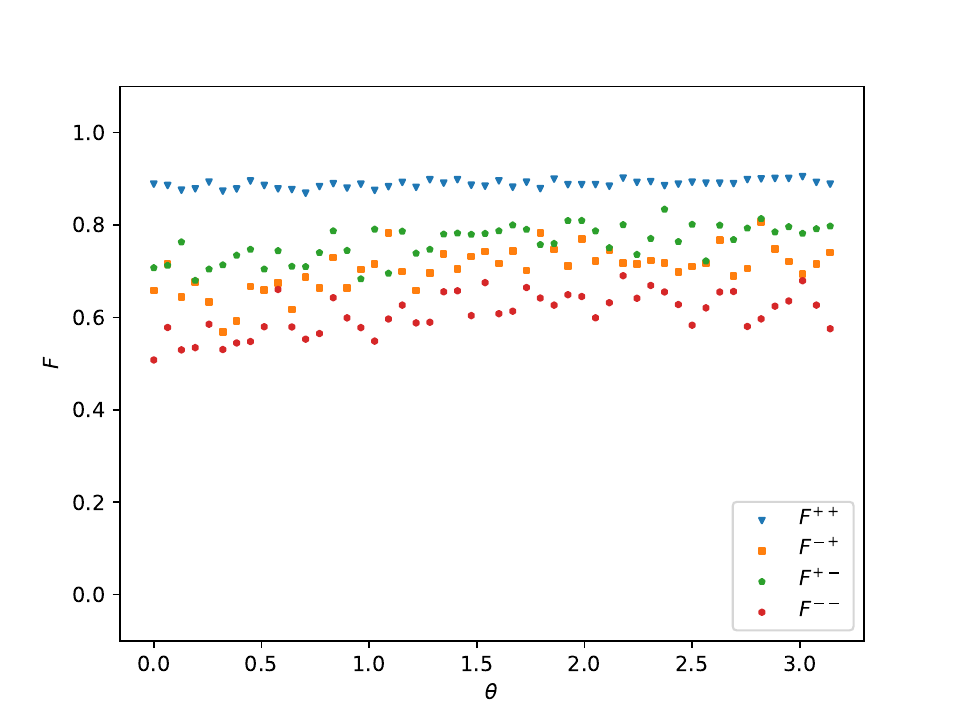}
                    \label{subfig: d_ico_fidelity_noise}
                }
                \caption{
                    The quantities $\braket{\hat{\sigma}_{\textrm{c}}^{x}}_{\pm}$, $\braket{\sigma_a^x}_{\pm}$, and $F^{\pm}$ measured from auxiliary circuits where perturbation absents is show in~\ref{sub@subfig: d_ico_p_noise} and \ref{sub@subfig: d_ico_fidelity_noise}, respectively.
                    These parameters are help to predict the theoretical results of measurement outcome in original circuits under the noise of experiments, which have shown in Fig.~\ref{fig: exp_ico_d}.
                }
                \label{fig: d_ico_exp_noise}
            \end{figure*}

            Denoting $F^{\pm \pm} \equiv \mathrm{Tr}\left[\rho_{\textrm{t}} \mathcal{E}^{\pm \pm} (\rho_{\textrm{t}})\right]$, where $\mathcal{E}^{\pm \pm} \equiv \mathcal{E}_{\textrm{a}}^{\pm} \circ \mathcal{E}_{\textrm{c}}^{\pm}$, it can be measured from the fidelity of error channel on $q_{\textrm{t}}$ with an initial state $\ket{\pm}_{\textrm{c}} \otimes \ket{\pm}_{\textrm{a}}$ of $q_{\textrm{c}}, q_{\textrm{a}}$.
            Similarly, we have
            \begin{equation}
                \mathrm{Tr}[\rho_{\textrm{t}} \mathcal{E}^{\pm \pm} (\hat{\sigma}^{\bm{r}}\rho_{\textrm{t}} \hat{\sigma}^{\bm{r}})] = r_z^2 F^{\pm \pm} + r_x^2 (1 - F^{\pm \pm}),
            \end{equation}
            and
            \begin{align}
                \mathrm{Tr}[\rho_{\textrm{t}} \mathcal{E}^{\pm \pm}(\hat{\sigma}_{\textrm{t}}^{\bm{r}} \rho_{\textrm{t}} \hat{\sigma}_{\textrm{t}}^{\bm{r}'})]
                = & \ \braket{0| \mathcal{E}^{\pm \pm}([\hat{\sigma}_{\textrm{t}}^{\bm{r}} |0\rangle _{\textrm{t}}\langle0| \hat{\sigma}_{\textrm{t}}^{\bm{r}'}])|0} \nonumber\\
                =& \sum_{i,j = 0,1} \braket{i| \hat{\sigma}_{\textrm{t}}^{\bm{r}} |0} \braket{0| \hat{\sigma}_{\textrm{t}}^{\bm{r}'} |j} \braket{0| \mathcal{E}^{\pm \pm}(|i\rangle _{\textrm{t}}\langle j|)|0} \nonumber\\
                = & \ \braket{0| \hat{\sigma}_{\textrm{t}}^{\bm{r}} |0} \braket{0| \hat{\sigma}_{\textrm{t}}^{\bm{r}'} |0} \braket{0| \mathcal{E}^{\pm \pm}(|0\rangle _{\textrm{t}}\langle0|)|0}
                + \braket{1| \hat{\sigma}_{\textrm{t}}^{\bm{r}} |0} \braket{0| \hat{\sigma}_{\textrm{t}}^{\bm{r}'} |1} \braket{0| \mathcal{E}^{\pm \pm}(|1\rangle _{\textrm{t}}\langle1|)|0} \nonumber\\
                & \ + \braket{0| \hat{\sigma}_{\textrm{t}}^{\bm{r}} |0} \braket{0| \hat{\sigma}_{\textrm{t}}^{\bm{r}'} |1} \braket{0| \mathcal{E}^{\pm \pm}(|0\rangle _{\textrm{t}}\langle1|)|0}
                + \braket{1| \hat{\sigma}_{\textrm{t}}^{\bm{r}} |0} \braket{0| \hat{\sigma}_{\textrm{t}}^{\bm{r}'} |0} \braket{0| \mathcal{E}^{\pm \pm}(|1\rangle _{\textrm{t}}\langle0|)|0}.
            \end{align}
            With $\bm{r} = (r_x, 0, r_z), \bm{r}' = (r_x, 0, - r_z)$,
            we have $\braket{0| \hat{\sigma}_{\textrm{t}}^{\bm{r}} |0} = - \braket{0| \hat{\sigma}_{\textrm{t}}^{\bm{r}} |0} = r_z, \braket{0| \hat{\sigma}_{\textrm{t}}^{\bm{r}} |1} = \braket{1| \hat{\sigma}_{\textrm{t}}^{\bm{r}} |0} = \braket{0| \hat{\sigma}_{\textrm{t}}^{{\bm{r}}'} |1} = \braket{1| \hat{\sigma}_{\textrm{t}}^{{\bm{r}}'} |0} = r_x$, thus
            \begin{align}
                \mathrm{Tr}[\rho_{\textrm{t}} \mathcal{E}^{\pm \pm}(\hat{\sigma}_{\textrm{t}}^{\bm{r}} \rho_{\textrm{t}} \hat{\sigma}_{\textrm{t}}^{{\bm{r}}'})]
                & = - r_z^2 F^{\pm \pm} + r_x^2 (1 - F^{\pm \pm}) + \imath r_x r_z \mathrm{Tr}[\hat{\sigma}_{\textrm{t}}^x \mathcal{E}^{\pm \pm}(|0\rangle _{\textrm{t}}\langle0|)] = r_x^2 - F^{\pm \pm}  .
            \end{align}

            The fidelities of ${\rho}_{\textrm{t}0}$ and ${\rho}_{\textrm{t}1}$ are
            \begin{align}
                F_{0}^{\pm} = & \ \frac{1}{12} \left[(1 + \braket{\hat{\sigma}_{\textrm{c}}^{x}}_{+})(1 + F^{+ \pm}) + (1 + \braket{\hat{\sigma}_{\textrm{c}}^{x}}_{-})(1 + F^{- \pm}) \right] \nonumber\\
                & \ + \frac{2 - r_x^2}{12} [(1 + \braket{\hat{\sigma}_{\textrm{c}}^{x}}_{+}) F^{+ \pm} - (1 + \braket{\hat{\sigma}_{\textrm{c}}^{x}}_{-}) F^{- \pm}]
                + \frac{r_x^2}{24} (\braket{\hat{\sigma}_{\textrm{c}}^{x}}_{+} - \braket{\hat{\sigma}_{\textrm{c}}^{x}}_{-}), \\
                F_{1}^{\pm}  = & \ \frac{2 - r_x^2}{12} [(1 + \braket{\hat{\sigma}_{\textrm{c}}^{x}}_{+}) F^{+ \pm} + (1 + \braket{\hat{\sigma}_{\textrm{c}}^{x}}_{-}) F^{- \pm}]
                + \frac{(2 r_x^2 - 1)}{12} (2 + \braket{\hat{\sigma}_{\textrm{c}}^{x}}_{+} + \braket{\hat{\sigma}_{\textrm{c}}^{x}}_{-}) \nonumber \\
                & \ + \frac{1}{12} [(1 + \braket{\hat{\sigma}_{\textrm{c}}^{x}}_{+}) F^{+ \pm} - (1 + \braket{\hat{\sigma}_{\textrm{c}}^{x}}_{-}) F^{- \pm}]
                + \frac{r_x^2}{24} (\braket{\hat{\sigma}_{\textrm{c}}^{x}}_{+} - \braket{\hat{\sigma}_{\textrm{c}}^{x}}_{-}), 
            \end{align}
            consequently, the fidelity of twirling channel with error is
            \begin{align}
                p_{++} = & \ \frac{2 + \braket{\hat{\sigma}_{\textrm{a}}^{x}}_{+} + \braket{\hat{\sigma}_{\textrm{a}}^{x}}_{-}}{16} \left[2 + \braket{\hat{\sigma}_{\textrm{c}}^{x}}_{+} + \braket{\hat{\sigma}_{\textrm{c}}^{x}}_{-} + \frac{2}{3} (\braket{\hat{\sigma}_{\textrm{c}}^{x}}_{+} - \braket{\hat{\sigma}_{\textrm{c}}^{x}}_{-})\right] \nonumber \\
                & \ + \frac{\braket{\hat{\sigma}_{\textrm{a}}^{x}}_{+} - \braket{\hat{\sigma}_{\textrm{a}}^{x}}_{-}}{16} \left[(2 + \braket{\hat{\sigma}_{\textrm{c}}^{x}}_{+} + \braket{\hat{\sigma}_{\textrm{c}}^{x}}_{-}) r_x^2 + \frac{1}{3} (r_x^2 + 1) (\braket{\hat{\sigma}_{\textrm{c}}^{x}}_{+} - \braket{\hat{\sigma}_{\textrm{c}}^{x}}_{-})\right], \\
                F^{e}_+ = & \ \frac{1}{4 p_{++}} \left[(1 + \braket{\hat{\sigma}_{\textrm{a}}^{x}}_{+}) F_{0}^{+}  + (1 + \braket{\hat{\sigma}_{\textrm{a}}^{x}}_{-}) F_{0}^{-} + (1 + \braket{\hat{\sigma}_{\textrm{a}}^{x}}_{+}) F_{1}^{+} - (1 + \braket{\hat{\sigma}_{\textrm{a}}^{x}}_{-}) F_{1}^{-}\right],
            \end{align}
            where $\braket{\hat{\sigma}_{\textrm{c}}^{x}}_{\pm}$, $\braket{\hat{\sigma}_{\textrm{a}}^{x}}_{\pm}$, and $ F^{\pm \pm}$ are measured from circuit of the twirling channel $\Lambda_{\mathrm{twirl}}^0$ of identical channel $\mathcal{I}$ with corresponding initial states $\ket{\pm}_{\textrm{c}} \otimes \ket{\pm}_{\textrm{a}}$ of $q_{\textrm{c}},q_{\textrm{a}}$ by post-selecting the state of $q_{\textrm{t}}$.
            The quantities $\braket{\hat{\sigma}_{\textrm{c}}^{x}}_{\pm}$, $\braket{\hat{\sigma}_{\textrm{a}}^{x}}_{\pm}$, and $F^{\pm\pm}$ measured from experiments is shown in Fig.~\ref{fig: d_ico_exp_noise}.

\end{widetext}


%

\end{document}